\documentclass[twocolumn]{aastex62}

\usepackage{graphicx}
\submitjournal{ApJ}

\shorttitle{Galaxy lookback evolution models}
\shortauthors{Kudritzki et al.}

\begin{document}


\title{Galaxy Lookback Evolution Models - a Comparison with Magneticum
  Cosmological Simulations and Observations }

\correspondingauthor{Rolf-Peter Kudritzki}
\email{kud@ifa.hawaii.edu}

\author{Rolf-Peter Kudritzki}
\affiliation{LMU M\"unchen, Universit\"atssternwarte, Scheinerstr. 1, 81679 M\"unchen, Germany}
\affiliation{Institute for Astronomy, University of Hawaii at Manoa, 2680 Woodlawn Drive, Honolulu, HI 96822, USA}
\author{Adelheid F.  Teklu}
\affiliation{LMU M\"unchen, Universit\"atssternwarte, Scheinerstr. 1,
  81679 M\"unchen, Germany}
\affiliation{Excellence Cluster Origins, Boltzmannstr. 2, 85748
  Garching, Germany}
\author{Felix Schulze}
\affiliation{LMU M\"unchen, Universit\"atssternwarte, Scheinerstr. 1, 81679 M\"unchen, Germany}
\affiliation{Max Planck Institute for Extraterrestrial Physics, Giessenbachstr. 1, 85748 Garching, Germany}
\author{Rhea-Silvia Remus}
\affiliation{LMU M\"unchen, Universit\"atssternwarte, Scheinerstr. 1, 81679 M\"unchen, Germany}
\author{Klaus Dolag}
\affiliation{LMU M\"unchen, Universit\"atssternwarte, Scheinerstr. 1, 81679 M\"unchen, Germany}
\affiliation{Max Planck Institute for Astrophysics,
  Karl-Schwarzschildstr. 1, 85748 Garching, Germany}
\author{Andreas Burkert}
\affiliation{LMU M\"unchen, Universit\"atssternwarte, Scheinerstr. 1,
  81679 M\"unchen, Germany}
\affiliation{Max Planck Institute for Extraterrestrial Physics, Giessenbachstr. 1, 85748 Garching, Germany}
\author{H. Jabran Zahid}
\affiliation{Microsoft Research, 14820 NE 36th St, Redmond, WA 98052, USA}
\

\begin{abstract}
We construct empirical models of star-forming galaxy evolution
assuming  that individual galaxies evolve along well-known scaling
relations between stellar mass, gas mass and star formation rate
following a simple description of chemical evolution. We test these models by a
comparison with observations and with detailed Magneticum high
resolution hydrodynamic cosmological simulations. Galaxy star
formation rates, stellar masses, gas masses, ages, interstellar medium
and stellar metallicities are compared. It is found that these simple
lookback models capture many of the crucial aspects of galaxy
evolution reasonably well. Their key assumption of a redshift
dependent power law relationship between galaxy interstellar
medium gas mass and stellar mass is in agreement with the outcome of
the complex Magneticum simulations. Star formation rates decline
towards lower redshift not because galaxies are running out of gas, but
because the fraction of the cold ISM gas, which is capable of producing
stars, becomes significantly smaller. Gas accretion rates in both model approaches are of
the same order of magnitude. Metallicity in the Magneticum simulations
increases with ratio of stellar mass to gas mass as predicted by the
lookback models. The mass metallicity relationships agree and the star
formation rate dependence of these relationships is also
reproduced. We conclude that these simple models provide a powerful
tool for constraining and interpreting more complex models based on
cosmological simulations and for population synthesis studies
analyzing integrated spectra of stellar populations.
  
\end{abstract}


\section{Introduction}

Observations of galaxies through cosmic time and detailed hydrodynamic
cosmological simulations show that the formation and evolution of
galaxies is an extremely complicated process. At high redshifts, the first building
blocks of galaxies contract and form the first stars while gas
continues to accrete from the intergalactic medium, providing new additional fuel for star formation. At the same time stars
produce heavy elements (metals)  through the nuclear fusion processes
in their interior. Some of the newly produced metals together with
some hydrogen and helium are recycled to the interstellar medium (ISM)
by a variety of complex stellar mass-loss processes. While stars
continue to form and gas is continuously accreted, metals accumulate
during the life of a galaxy, but at the same time a significant
fraction of the metals appears to be expelled from the ISM by large scale
galactic winds. In addition, merging processes with infalling other nearby
galaxies influence the evolution significantly. 

In view of the complexity of these many processes and their interplay
on different time scales it is surprising that an intriguingly simple
relationship exists between total galactic stellar mass and the average
metallicity of galaxies, the mass-metallicity relationship (``MZR''),
see for example \citet{Lequeux1979}, \citet{Tremonti2004},
\citet{Kudritzki2016}(herafter K16), \citet{Zahid2017} (herafter
Z17). This MZR and its evolution with redshift appears like a true Rosetta
stone to understand the key aspects of galaxy evolution. For instance,
\citet{Zahid2014} (hereafter Z14) show that the observed
MZRs at different redshift can be explained by a very simple model with galactic winds and accretion
where the observed metallicity is a function of the ratio of
galactic stellar to ISM gas mass. Over their lifetime star forming galaxies evolve along the (redshift
dependent) main sequence of star formation and turn gas into
stars. During this evolution
the low-mass metal-poor galaxies are gas-rich and the high-mass metal-rich galaxies are
gas-poor (see Fig. 5 and 7 in Z14).

Given the success of the Z14 approach in matching and explaining the
observations it appears important to further investigate the validity of this
rather simple galaxy evolution model. An obvious way is the comparison
with cosmological simulations, which describe the complicated
processes during the formation and evolution of galaxies in a much
more comprehensive way. This is done in the following by using the extensive
Magneticum simulations (see section 4).

Starting from the ideas described in Z14 and Z17 we develop a new
generation of lookback models, which describe the evolution of
galaxies. We then compare the properties of these models (SFR
$\psi$, stellar mass M$_*$,  ISM gas mass M$_g$, luminosity weighted age t
of the stellar population and logarithmic metallicity [Z]
of the ISM and the stellar population) with the properties of galaxies in the
Magneticum Box 4 (high resolution) simulations. We also compare with observations.

\section{Lookback models}

Our goal is to describe the evolution of a galaxy, which is observed
at a redshift z$_0$ with a stellar mass M$^*$(z$_0$) back to
its origin. Because we are looking back in cosmic time, we call these
models lookback models. The relationship between lookback time and
redshift z is given by the standard equations

\begin{equation}
  t(z) = t(z_0) + {1 \over H_0}\int^z_{z_0}{dz \over (1+z)E(z)}
\end{equation}

with E(z)

\begin{equation}
  E(z) = (\Omega_{\Lambda} + \Omega_m(1+z)^3)^{1 \over 2}
\end{equation}

and

\begin{equation}
  t(z_0) =  {1 \over H_0}\int^{z_0}_{0}{dz \over (1+z)E(z)}.
\end{equation}

Like the Magneticum simulations (see section 4) we adopt a flat universe with H$_0$ =
70.4 km/sec/Mpc = 70.4$\times$1.023$^{-12}$yr$^{-1}$ and densities $\Omega_{\Lambda}$ = 0.728 and
$\Omega_m$ = 1 - $\Omega_{\Lambda}$.

The mass evolution is then described by

\begin{equation}
 M_*(z) = M_*(z_0) -{(1-R) \over H_0}\int^z_{z_0} {\psi(M_*,z) \over (1+z)E(z)}dz.
 \end{equation} 

R is the fraction of stellar mass which is returned to the
interstellar medium because of stellar winds and supernova
explosions. Following Z14 and Z17 we adopt R = 0.45 based on the
assumption of a \citet{Chabrier2003} stellar initial mass function. $\psi$ is the
star formation rate (SFR) as a function of stellar mass and
redshift. We use a modified form (see section 5 below)  of the SFR law
from \citet{Pearson2018}, Appendix C, which is based on observations of galaxies
on the star foming main sequences out to redshift z = 6. We usually
finish the integration, when a minimum stellar mass of
10$^6$ M$_{\odot}$ is reached.

A key simplification in our approach is the assumption that at every
redshift there is a power law correlation between the total mass of
the cold (molecular and atomic) gas of the ISM and the stellar mass of galaxies

\begin{equation}
  M_g(z) = A(z)M_*^{\beta}(z)
  \end{equation}

  with

  \begin{equation}
A(z) = A_0(1+z)^{\alpha}.
  \end{equation}

This assumption  is based on survey
observations of star forming galaxies in the local Universe, which
indicate a power law holding over several orders of magnitude in
stellar mass with an exponent
of the order of $\beta$ $\sim$ 0.5 (see Z14, \citealt{Peeples2014},
\citealt{Saintonge2017}, \citealt{Catinella2018}, \citealt{Hunt2020}). The redshift
dependence was introduced by Z14 to match the observed ISM
oxygen abundance MZRs out to higher redshift. We will use
the same procedure for our new models and give new values for
$\alpha$, $\beta$ and A$_0$ below.

With the use of equations (4) and (5) we do not account for
  potential strong variations of star formation activity caused by a
  multitude of processes, during which active galaxies become passive
  or passive galaxies are rejuvenated to become active again (see, for
  instance, \citealt{Trussler2020, Spitoni2020}). The concept of our
  lookback models is that despite of this deficiency they still
  sufficiently well describe the average evolution and physical
  properties of galaxies observed on the main sequences at different
  redshift. The comparison with the detailed Magneticum simulations,
  which include variations of star formation activity (see below), can
  then be used to assess whether this concept is valid.

The chemical evolution is described by the metallicity equation for
the metallicity mass fraction Z (for all details see Z14)

\begin{equation}
  {dZ \over dM_*} = {1 \over M_g}({dM_Z \over dM_*} - Z{dM_g \over dM_*}),
\end{equation}

where the change of the metallicity mass M$_Z$ of the ISM is given by

\begin{equation}
{dM_Z \over dM_*} = {Y_N \over 1-R} - Z.
\end{equation}

 Y$_N$ is the effective yield,  which in addition to the stellar
 nucleosynthesis yield Y includes the effects of accretion
 from the halo and the intergalactic medium (dM$_{\mathrm{accr}}$) and galactic
 winds (dM$_{\mathrm{wind}}$) as
 described by Z14

 \begin{equation}
Y_N = Y -\zeta
\end{equation}

with

\begin{equation}
\zeta = \left(Z_{\mathrm{wind}}{dM_{\mathrm{wind}} \over dM_*} - Z_{\mathrm{accr}}{dM_{\mathrm{accr}} \over dM_*}\right)(1-R).
\end{equation}.

 Z$_{\mathrm{accr}}$ is the metallicity of the accreted gas, while
 Z$_{\mathrm{wind}}$ corresponds to the metallicity of the matter lost
through galactic winds. Based on observational evidence provided by
\citet{Zahid2012} and \citet{Peeples2014},  Z14 argue convincingly that $\zeta$$\sim$
const. is a reasonable approximation. We therefore use Y$_N \sim$
const. as a free parameter noting that it consists of two components,
the nucleosynthesis yield Y and the effects of accretion and winds
described by $\zeta$.

Using eq. (5) and (6) we can express ${dM_g \over dM_*}$ as

\begin{equation}
{dM_g \over dM_*} = \beta{M_g \over M_*}(1 - K(M_*,z))
\end{equation}

with
\begin{equation}
K(M_*,z) = {\alpha \over \beta}H_0{E(z) \over (1-R)\psi}M_* .
\end{equation}

The function K describes the influence of the evolution with redshift
of the power law relation between stellar mass and gas mass.  

Eq. (11) then leads to the final form of the metallicity equation

\begin{equation}
 {dZ \over dM_*} = {1 \over M_g}\{ {Y_N \over 1 - R} - Z\left (1+\beta{M_g \over M_*}(1-K) \right) \}.
\end{equation}

 We note that Z14 in their analytical approach neglect the second term
 on the right hand side of eq. (7). However, solving our set of
 lookback model equations numerically and calculating a large grid of
 models with different model parameters we find that including this term
 leads to a small, but non-negligible quantitative difference in the calculated
 metallicity during the evolution of a galaxy. We, therefore, keep
 this term for the calculation of our numerical lookback models
 and develop a new analytical solution later in Appendix A.

\begin{figure}[ht!]
 \begin{center}
  \includegraphics[width=0.40\textwidth]{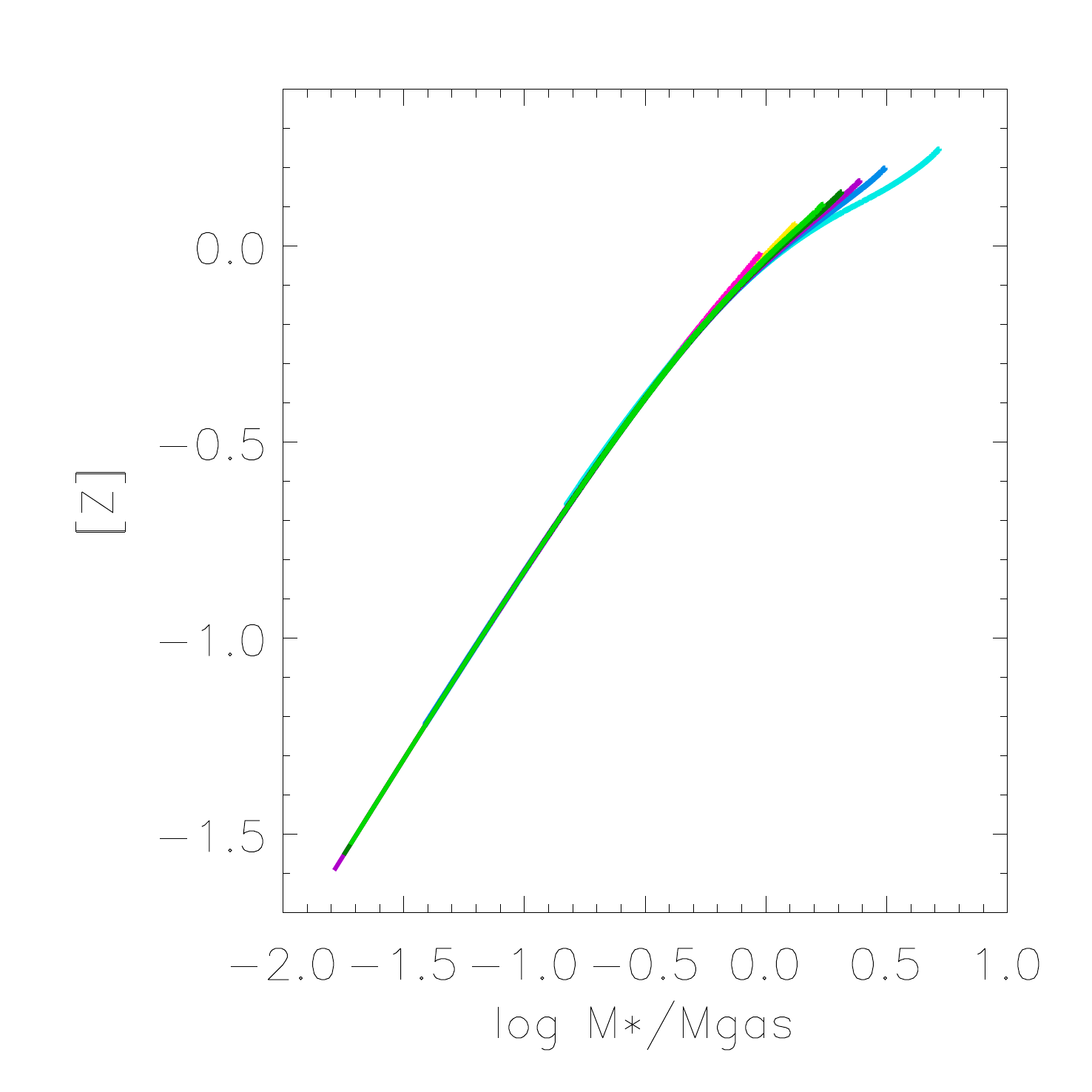}
  \caption{
Lookback model ISM metallicity [Z] as a function of the ratio of stellar mass to gas mass for the evolution of 7
galaxies with final masses of log M$_*$ = 9.28, 9.65, 10.01, 10.16, 10.34, 10.60,
11.13. We use different colors for the galaxy evolution tracks with different
final mass. Note that the tracks lie on top of each other, which means
that [Z] depends only on M$_*$/M$_g$. For the defintion of [Z] see
equation (14). The result is discussed in detail in Appendix A and 
the calculations are described in section 2. \label{Fig1} }
 \end{center}
\end{figure}
 
 Fig.~\ref{Fig1} shows a typical result for the lookback model
 evolution of seven galaxies with different final masses ($\alpha$ =
 0.4, $\beta = 0.6$, log A$_0$ = 3.73, log${Y_N \over
   Z_{\odot}(1-R)}$ = 0.2 and the star formation law described in
 section 5 have been used
 for these calculations). While there is a small
 quantitative difference to the Z14 results, we see that the most
 important property of the lookback models remains. Metallicity is to
 a good approximation solely a function of the ratio of stellar mass
 to gas mass M$_*$/M$_g$. As we show in Appendix A, this is the consequence of the key
 assumption of our lookback models, the relationship between gas mass
 and stellar mass as described by eq. (5) and (6). The comparison with
 the Magneticum models will be an important check whether this key
 assumption is valid.

 In the following we will express metallicities in units of the solar
 metallicity defined as
 
 \begin{equation}
   [Z] = {\rm log}~Z/Z_{\odot}.
 \end{equation}
  
For the metallicity mass fraction of the sun we will use  Z$_{\odot}$
= 0.014 \citep{Asplund2009}.

The metallicity shown in Fig.~\ref{Fig1} is the metallicity of the
ISM and the stars just born from this ISM. For the comparison with stellar metallicities we will also
calculate a V-band luminosity weighted average over the whole galactic
population of stars, which includes all stars that formed earlier at lower
metallicity. Details of the calclulations are described in Appendix C.

At the end of this section it is appropriate to briefly
  compare our lookback models with most recent alternative analytical
  models describing galaxy evolution. \citet{Spitoni2017, Spitoni2020}
  present a model with an exponentially decreasing gas infall rate, a
  star formation rate proportional to total cold gas mass and a gas
  outflow rate proportional to star formation rate.  They do not consider the fact that only
  the molecular fraction of the cold gas is involved in the star
  formation process, which may be important in the light of section
  6.5 further below. Gas accretion time scale, total accreted gas mass and mass loading
  factor are free parameters.  With assumptions about star formation time and a power law relationship between gas mass and stellar mass (both redshift independent) they use observations of the MZR and the SFR main sequence at z=0.1 and 2.2 to constrain accretion time scales, ages, mass loading factors and accreted gas masses of star forming and passive galaxies and to investigate the effects of galaxy downsizing.

  \citet{Pantoni2019} make similar assumptions about the time dependence
of infall and star formation rate but they relate the amount of
infalling mass to the baryonic mass present in the host halo, which is
characterized by a dark matter mass distribution following an
NFW-profile \citep{Navarro1997}.
The accretion time scale is obtained from estimates of the
halo cooling and dynamical time scales which depend on assumptions about
density and clumping. The star formation time scale is obtained from a
redshift independent estimate of the dynamical time scale in the
rotating galactic disk at the radius of centrifugal equilibrium. Gas
outflow rates are adopted from theoretical work assuming energy or
momentum driven winds. For the estimate of halo masses, mass growth by
merging processes is taken into account using fitting formulae
obtained from the comparison with cosmological simulations. After
constraining all parameters \citet{Pantoni2019} apply their models to
the star forming progenitors of early type galaxies and calculate SFR,
gas mass and metallicities as a function of stellar mass at high
redshifts from z = 2, 4 and 6. 
    
\citet{Lapi2020} extend the \citet{Pantoni2019} model by
  adding the effects of wind recycling and galactic fountains. They
  also consider the important effect that star formation is related to
  the molecular fraction of interstellar gas only and they modify star
  formation time accordingly. They calculate the fraction of the star
  forming to the total cold ISM gas assuming a linear dependence on
  the mass weighted Toomre parameter. As a result the ratio of mass
  accretion time to star formation time changes significantly compared
  to \citet{Pantoni2019}. The agreement with observations of star formation rates, gas and dust masses, metallicities and specific stellar angular momentum as a function of stellar mass at redshifts z = 0 and 1 of star formation rates is remarkably good. This is a self-consistent approach, which aims to account for the multitude of individual processes through a variety of parameters which are constrained mostly by theoretical arguments and comparisons with numerical simulations.

In contrast, our lookback models do not consider the many different individual processes affecting galaxy evolution. They avoid explicit assumptions about mass accretion and mass loss as a function of time and instead condense the complex interplay between accretion, outflows and star formation into the simple redshift dependent relationship between gas mass  and stellar mass. As a result, the understanding of chemical evolution is straightforward and related to the effective yield and the ratio of stellar mass to gas mass. As we will see in the following, this simple approach agrees well with observations and detailed hydrodynamical cosmological simulations.

\section{Observations}

To compare our model calculations with observations
we use galaxy metallicities derived from quantitative stellar spectroscopy of individual
blue and red supergiant stars in nearby galaxies out to 20 Mpc (see
K16) and population synthesis stellar spectroscopy of stacked
spectra of 250000 SDSS galaxies at a redshift of z $\sim$ 0.08 (see
Z17). We also include observational results obtained
from HII region strong line studies at z = 0.08, 0.29, 0.78, 1.55, 2.3
and 3.3 from \citet{Sanders2020} (violet circles: their Table 1;
yellow circles:  eq. 10 with SFR
from eq. 2, 3, 4), \citet{Genzel2015} (their eq. 12a) and Z14 (their Table 2
and eq. 5). The Z14 strong line oxygen abundances were based on the \citet{Kobulnicky2004} calibration. According to the work by K16 and
\citet{Bresolin2016} the N2 calibration by \citet{Pettini2004}
is more appropriate. We have therefore transformed the Z14
abundances to this calibration using the transformation suggested by
\citet{Kewley2008}. The \citet{Genzel2015} and \citet{Sanders2020}
oxygen abundances were already determined using the
\citet{Pettini2004} calibration and do not require a transformation adjustment.

\begin{figure}[ht!]
 \begin{center}
   \includegraphics[width=0.48\textwidth]{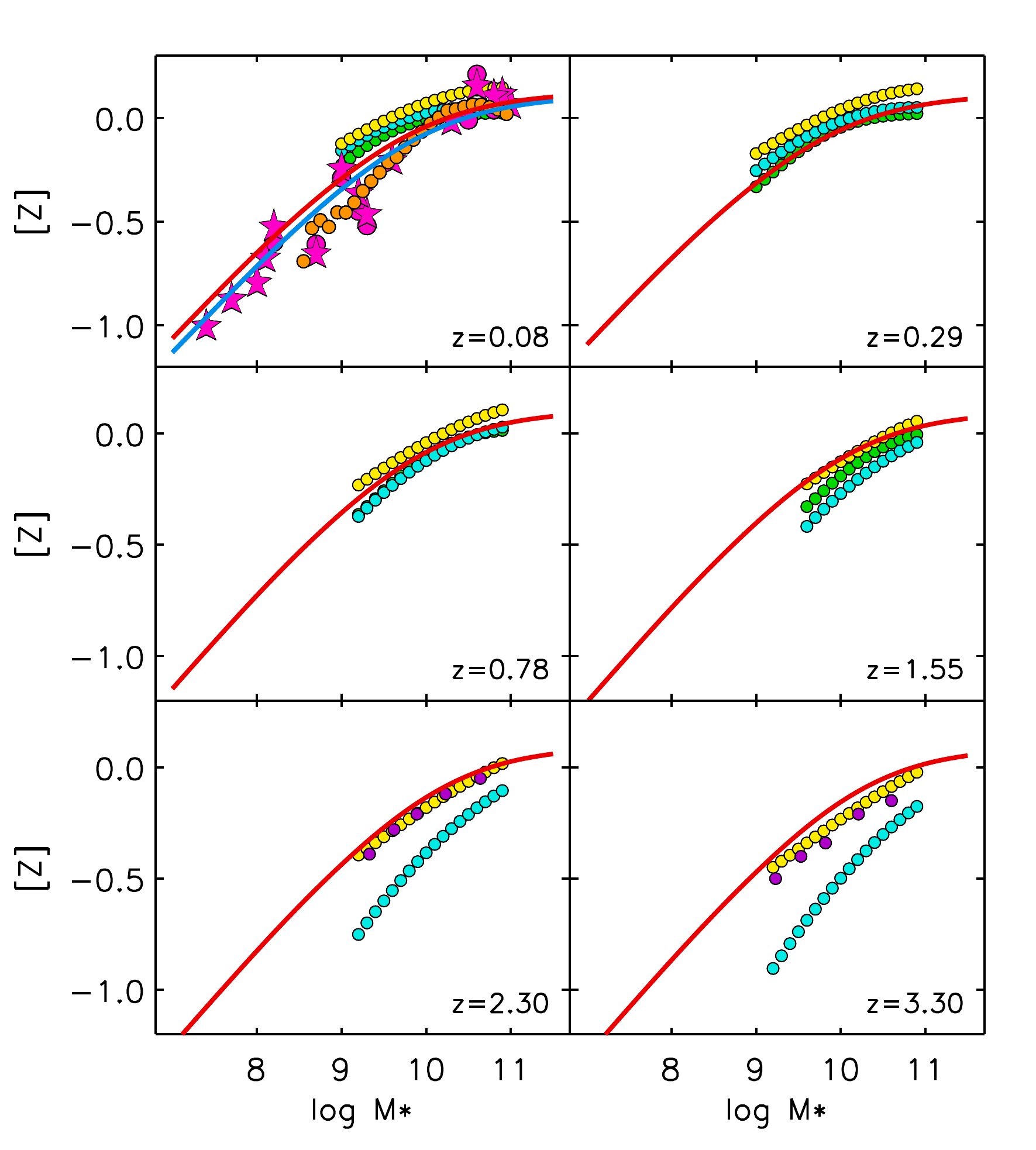}
  \caption{
Lookback model MZRs at different redshifts. The metallicity of the ISM
and the young stellar population versus the total stellar mass of the
galaxy is shown in red. The blue curve represents the V-band luminosity weighted
metallicity of the whole stellar population. Observed HII-region MZRs
at the same redshift are shown in green \citep{Zahid2014}, cyan
\citep{Genzel2015}, yellow and violet
\citep{Sanders2020}. For the lowest redshift we also show observations
of the stellar
metallicities of the young stellar population (pink circles: red
supergiant stars; pink stars: blue supergiant stars, see
\citealt{Kudritzki2016}). The orange
circles respresent metallicities of the young stellar population
obtained by population synthesis analysis of SDSS spectra of the
integrated stellar population of a large sample of galaxies (\citealt{Zahid2017}). 
\label{Fig2} }
 \end{center}
\end{figure}

Fig.~\ref{Fig2} shows the comparison of our lookback models
with the observations. The calculations use the parameters
$\alpha$ = 0.40, $\beta$ = 0.55, log A$_0$ = 4.28 (see eq. 5 and 6,
M$_*$ and M$_g$ in solar masses) and  log${Y_N \over Z_{\odot}(1-R)}$
= 0.100 and apply a main sequence star formation law as described in
section 5 with the correction factor c(z) set to unity. We find remarkable agreement.

\section{Magneticum data}

The Magneticum\footnote{www.magneticum.org} simulations are a set of fully hydrodynamical cosmological
simulations of different box-volumes and resolutions. They follow the formation and evolution of 
cosmological structures through cosmic time, accounting for the complex physical processes which
shape the first building blocks of galaxies into the mature galaxies of today. For details on these simulations 
see \citet{Hirschmann2014} and \citet{Teklu2015}. A WMAP-7 $\Lambda\mathrm{CDM}$  cosmology \citep{Komatsu2011}
is adopted with $h=0.704$, $\Omega_m = 0.272$, $\Omega_b = 0.0451$, $\Omega_\lambda = 0.728$, $\sigma_8 = 0.809$,
and an initial slope of the power spectrum of $n_s = 0.963$. 

Specifically, as this is relevant to the work presented here, metal radiative cooling is implemented
according to \citet{Wiersma2009}, i.e. the interstellar medium is treated as a two-phase medium 
where clouds of cold gas are embedded in the hot gas phase. Star
formation and galactic winds are treated in the same way as described
by \citet{Springel2003}. Metals and energy are released by stars 
of different mass by integrating the evolution of the stellar population \citep[for details see][]{Dolag2017}, 
properly accounting for mass-dependent lifetimes using a lifetime function according to \citet{Padovani1993},
the  metallicity-dependent stellar yields by \citet{Woosley1995} for SNe II, the yields by \citet{Hoek1997}
for AGB stars, and the yields by \citet{Thielemann2003} for SNeIa. Stars of different mass are
initially distributed according to a Chabrier initial mass function \citet{Chabrier2003}.

For our comparison with the model calculations and observations we use the Magneticum Box 4 uhr simulation.
This simulation has a Box volume of (48  Mpc/h)$^3$ with initially 2x5763 (dark matter and gas) particles.
The particle masses are $m_\mathrm{DM} = 3.6\times10^7M_\odot$/h and
$m_\mathrm{Gas} = 7.3\times10^6M_\odot$/h, respectively, and each gas particle can spawn up to four stellar  particles (i.e. the stellar particle mass is approximately  
1/4th of the gas particle mass), with a softening of $\epsilon_\mathrm{DM}=\epsilon_\mathrm{Gas} = 1.4$ kpc/h
and $\epsilon_*= 0.7$ kpc/h.

From this simulation we use two data sets: The first one
is a snapshot of star forming galaxies at different redshifts, z $\sim$ 0.1
(exact 0.07), $\sim$ 0.5 (0.47), $\sim$ 1.0 (0.99), $\sim$ 1.5 (1.48),
$\sim$ 2.0 (1.98), $\sim$ 2.8 (2.79), $\sim$ 3.4 (3.42), $\sim$ 4.2
(4.23), $\sim$ 5.3 (5.34), $\sim$ 6.9 (6.94). The second data set combines galaxies in 
stellar mass bins at redshift z = 0.07 and traces their evolution back
in time at each time (or redshift) calculating averaged properties of
the sample. The mass bins are log M$_*$ = 9.28, 9.65, 10.01, 10.16, 10.34, 10.60,
11.31 and the evolution is traced up to redshift z = 4.2. The selected
mass bins are an arbirary choice to represent the range from lower to
higher galaxy stellar masses. We call the first data set the redshift ``snap shot sample''
and the second one the ``evolution sample''. In the snap shot sample
we distinguish between disk galaxies and intermediate galaxies. For
this we use the position of a galaxy in the stellar mass--angular momentum plane, quantified by the b-value

\begin{equation}
b = \mathrm{log}({j \over \mathrm{kpc~km/s}}) - {2 \over 3} \mathrm{log}({M_* \over M_{\odot}}),
\end{equation}

where j is the specific angular momentum of the galaxy stellar
component (see especially \citealt{Teklu2017}, but also
  \citealt{Teklu2015} and \citealt{Schulze2018} for more details).
At redshift z = 0.07, galaxies with -4.35 $\geq$ b $\geq$ -4.75 are classified as
intermediates, whereas galaxies with b  $\geq$ -4.35 are considered
disks. Note that intermediates are a transitional galaxy type between
disks and spheroids.

We emphasize that some of the selected galaxies in our snap
  shot sample went through earlier phases of strongly reduced star
  formation and were then rejuvenated again by gas rich merging
  processes. We keep these galaxies in our sample. As already
  discussed in section 2 this allows us to test, whether the lookback
  models, which are thought to describe the averaged evolution of star
  forming galaxies and do not account for strong changes of star
  formation, describe the average properties of star forming galaxies well.

For the evolution sample we select all disk galaxies at z = 0.07 with
log M$_* \ge$ 9. Note that at higher redshift these galaxies could
also have been intermediates or spheroids or also passive.

\section{Star formation rates}

\begin{figure}[ht!]
 \begin{center}
   \includegraphics[width=0.47\textwidth]{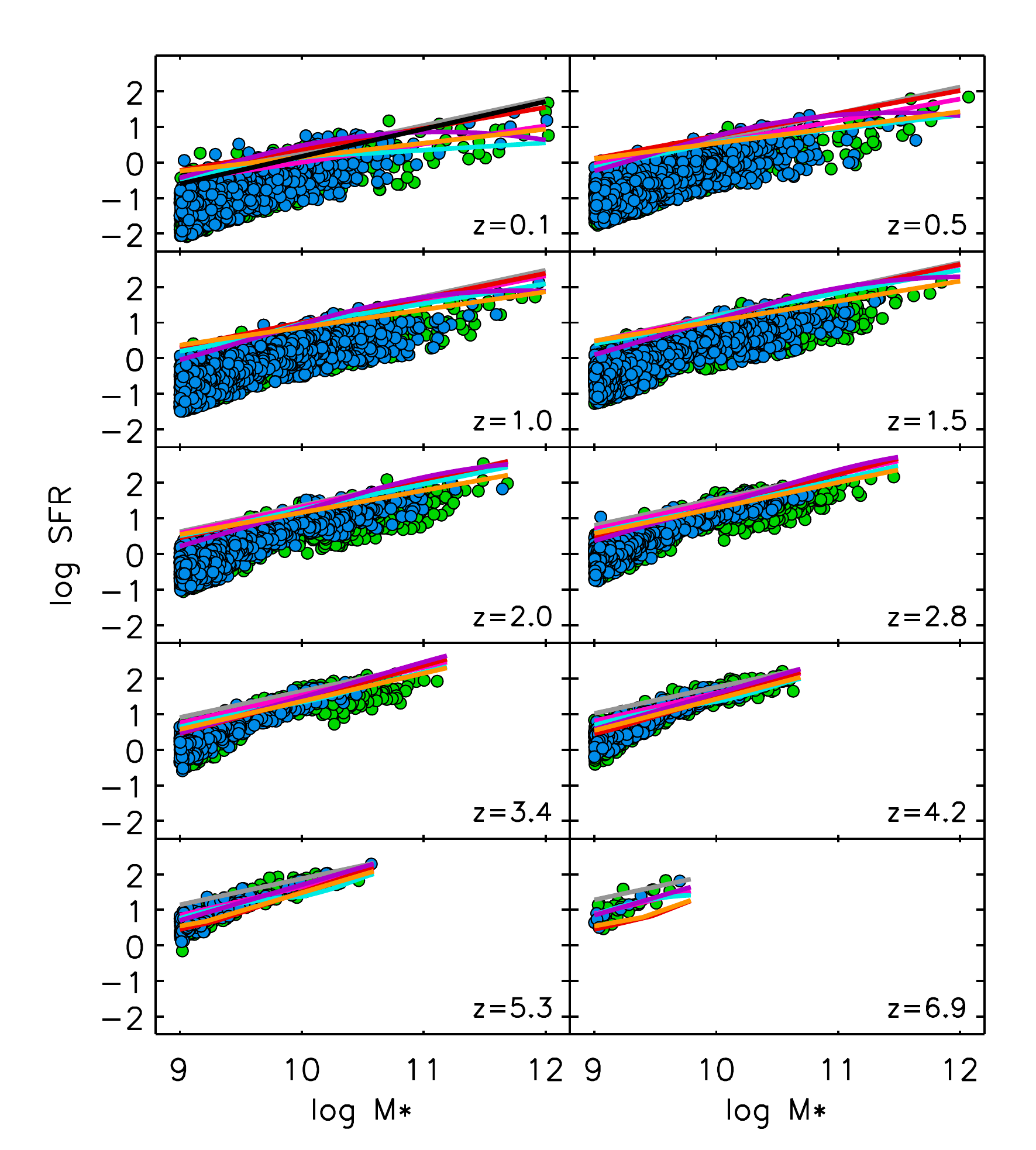}
  \caption{
Logarithm of star formation rates of the Magneticum snap shot samples at different
redshifts as a function of the logarithm of stellar mass (green
circles: intermediates, blue circles: disks). Observed star
formation rates are overplotted as solid curves: black: \citet{Elbaz2007},  cyan:
\citet{Behroozi2013}, violet: \citet{Schreiber2015}, orange:
\citet{Pearson2018}, red: \citet{Pearson2018}, Appendix C, pink:
\citet{Speagle2014}, grey: \citet{Sanders2020}.
   \label{Fig3} }
 \end{center}
\end{figure}

For the comparison with the Magneticum simulations we need to adopt a
SFR law $\psi$(M$_*$,z), which provides SFRs as a function of stellar
mass and redshift. There is a rich literature of observed SFR laws of the
so-called galaxy main sequences (see, for instance,
\citealt{Pearson2018}, \citealt{Tacconi2020} for a recent summary).  Most of the approaches describe star
formation as a power law

\begin{equation}
  \psi(z, M_*) = \psi_0(z)M_*^{\delta(z)},
 \end{equation}

 where both the zero point $\psi_0(z)$ and the slope $\delta(z)$ increase
 with redshift. $\psi$ is given in solar masses per year and the
 stellar masses are in solar units. We have selected a few published SFR laws
 (\citealt{Elbaz2007}, \citealt{Behroozi2013}, \citealt{Speagle2014}, 
 \citealt{Schreiber2015}, \citealt{Pearson2018}, \citealt{Sanders2020}) and compare them with
 the Magneticum SFRs in Fig.~\ref{Fig3}. Given the systematic
 uncertainties of the observations (see the example discussed below) the comparison is
 reasonable and indicates that the sub-grid physics treatment of star
 formation adapted from \citet{Springel2003} works well. However, we note a systematic difference at low and intermediate redshifts where the
 Magneticum SFRs are lower by a factor of 2 to 3.

\begin{figure}[ht!]
 \begin{center}
   \includegraphics[width=0.45\textwidth]{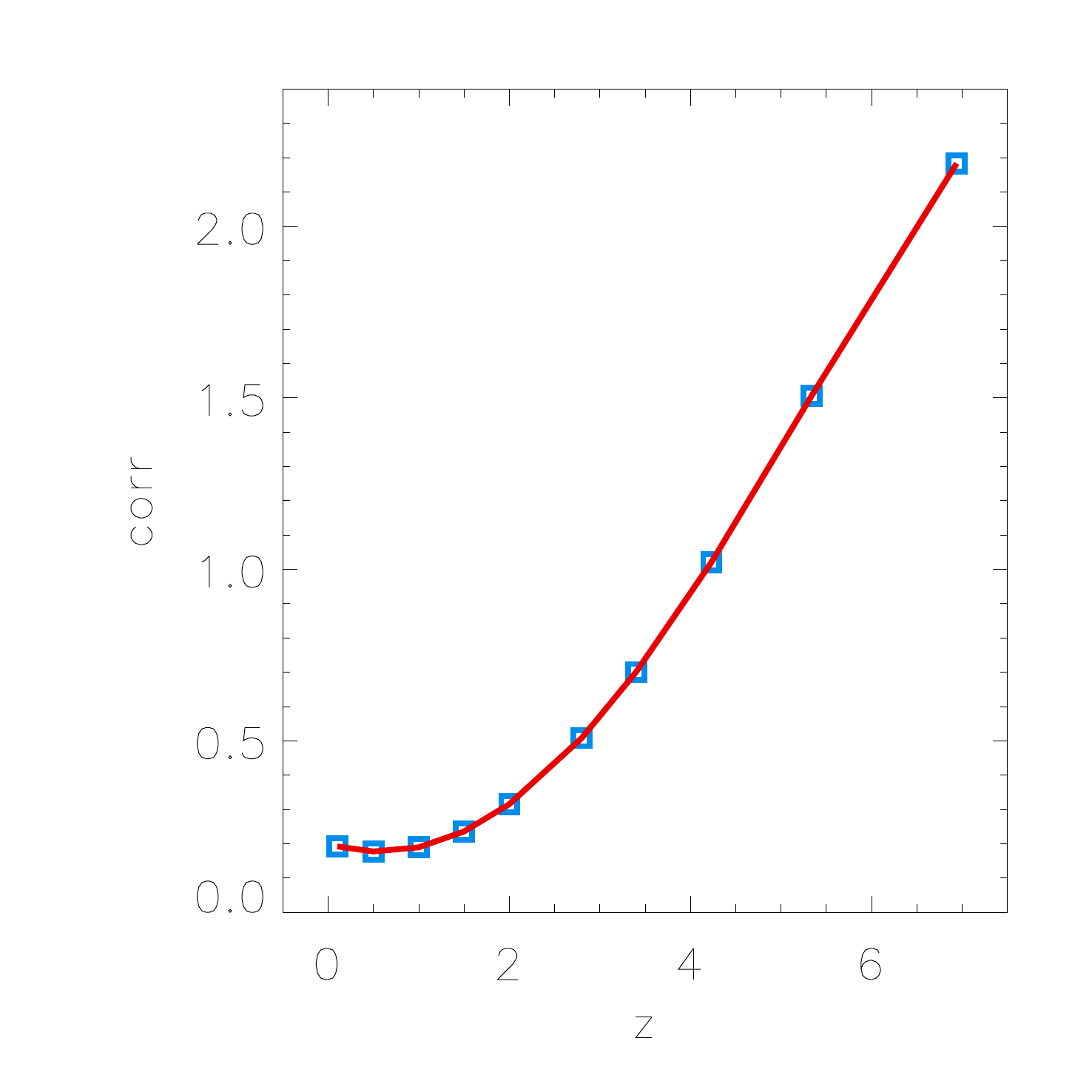}
  \caption{
Correction factor applied to the \citet{Pearson2018} star formation
rates as a function of redshift to match the Magneticum SFRs (see text).
   \label{Fig4} }
 \end{center}
\end{figure}
 
In principle, we can construct lookback models with any reasonable
star formation main sequence representation. For instance, for the
comparison with observations in Fig.~\ref{Fig2} we have used
\citet{Pearson2018} as described in their Appendix C. However, the
purpose of this work is to compare our simple lookback model approach
with the highly complex and detailed hydrodynamic Magneticum
simulations. Therefore, it is prudent to adopt a SFR description
which matches the Magneticum SFRs reasonably well. We accomplish this
by again using  the \citet{Pearson2018}
 SFR law of their Appendix C, however now with several
 modifications. Most importantly, we apply a redshift dependent correction factor c(z) to $\psi_0(z)$

\begin{equation}
  \psi_0^{LB}(z)  = \psi_0(z)c(z),
 \end{equation}

 which is shown in Fig.~\ref{Fig4}. With this factor we obtain a star
 formation law $\psi_{LB}(z,M_*)$ which is in
 good agreement with the Magneticum SFRs at M$_*$ =
 10$^{10.5}$M$_{\odot}$.  In addition, we introduce a broken power law with respect
 to stellar mass, which is described in more detail in Appendix B
 together with the complete description of $\psi_{LB}(z,M_*)$ .

 \begin{figure}[ht!]
 \begin{center}
   \includegraphics[width=0.45\textwidth]{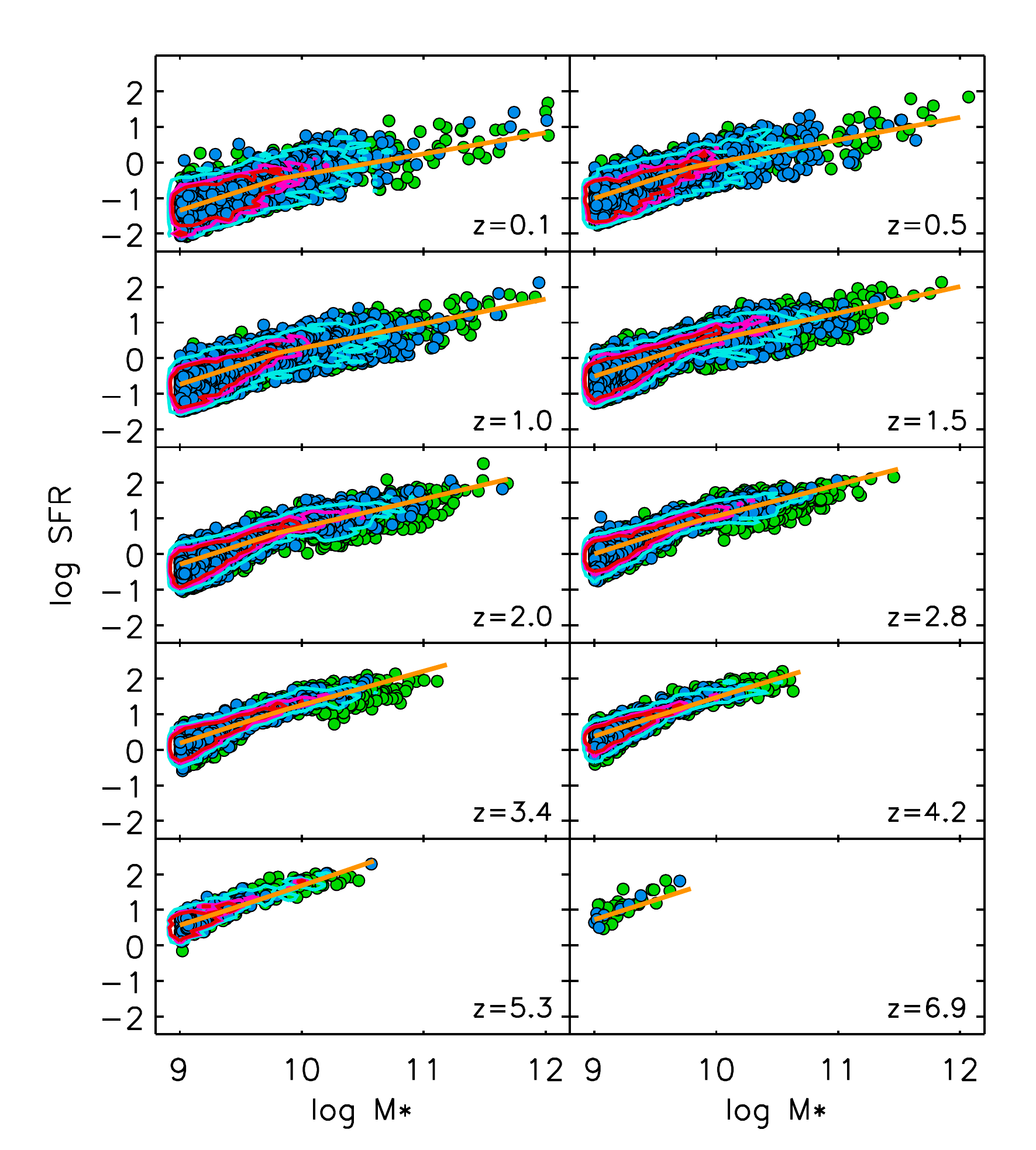}
 \caption{
Same as Fig.~\ref{Fig3} but now with the adopted lookback model star
formation rates overplotted in orange (see text). We also show
  isocontours enclosing 68\% (red), 80\% (pink) and 95\% (cyan) of the
  Magneticum galaxies.
  \label{Fig5} }
 \end{center}
\end{figure}

\begin{figure}[ht!]
 \begin{center}
   \includegraphics[width=0.45\textwidth]{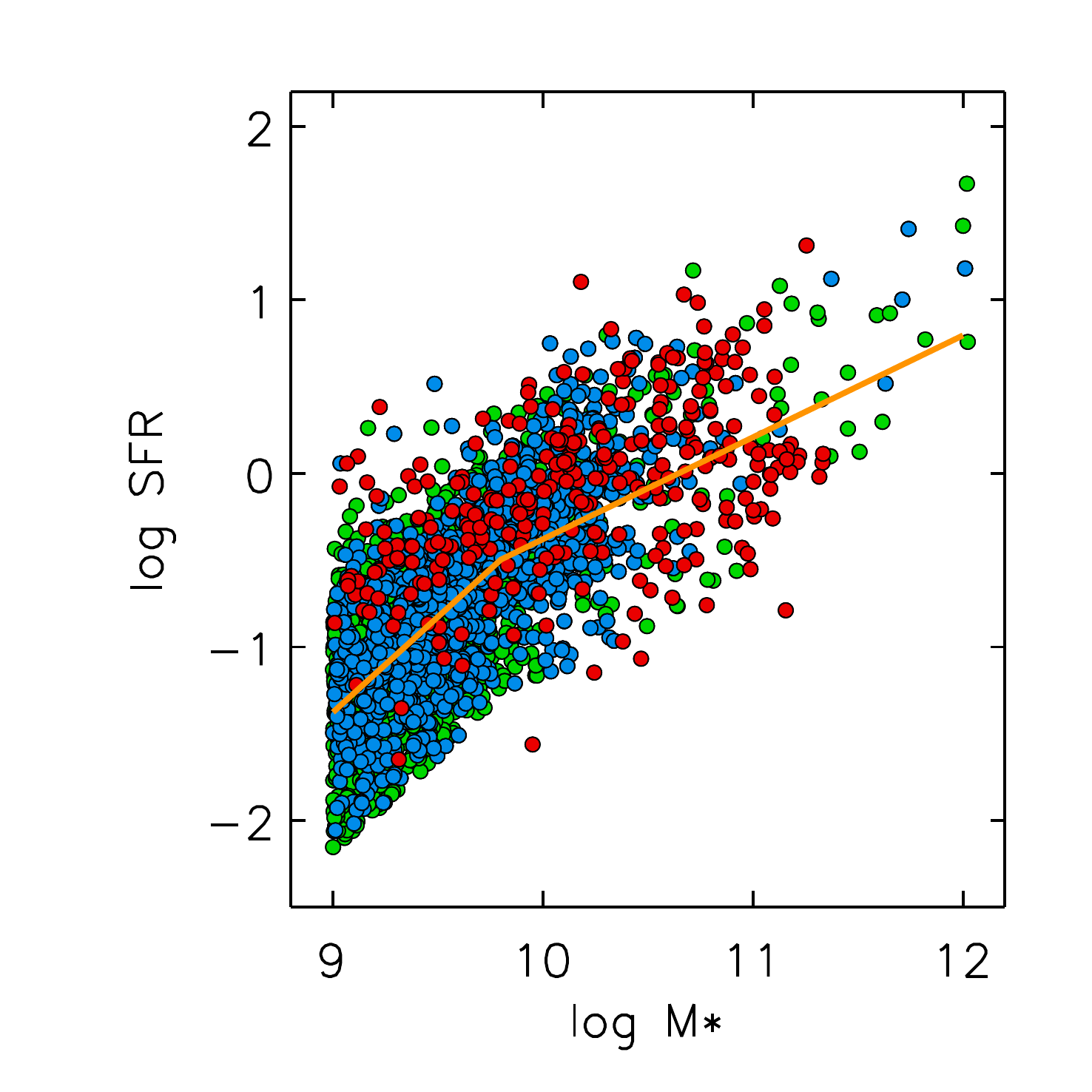}
   \caption{
Magneticum snap shot star formation rates at redshift z = 0.1 (green
and blue circles as before)  compared with the observed star formation
rates of the xGASS and xCOLDGASS surveys (red circles, for references
see text). The adopted lookback model star
formation rates are overplotted in orange.
  \label{Fig6} }
 \end{center}
\end{figure}

 Fig.~\ref{Fig5}  shows that $\psi_{LB}$ provides a good representation of the average
 SFRs of the Magneticum snap shot sample. While the literature main
 sequence SFRs shown in Fig.~\ref{Fig3} indicate that our adopted
 Magneticum matching star formation law $\psi_{LB}$ may underestimate
 the observed main sequence SFRs at lower and intermediate redshift,
 the comparison in Fig.~\ref{Fig6}  with observed SFRs obtained by the
 xGASS and xCOLDGASS surveys (\citealt{Saintonge2017}, \citealt{Catinella2018}) of
 nearby galaxies indicates good agreement. We also note that the MAGMA
 survey (see \citealt{Hunt2020}) agrees well with Magneticum and
 $\psi_{LB}$ except at the very low mass end of log M$_* \approx$ 9,
 where the MAGMA SFRs are 0.3 dex larger.  While the physical
 reason for the introduction of the correction
 factor c(z) may, thus, be caused by systematic uncertainties of the
 methods how star formation rates are determined, we also have to note
 a weakness of the simulations. They adopt a local star formation
 efficiency, which is constant over time for a given amount of cold
 gas density above a certain threshold, with only minor variations
 due to changes of the ISM cooling function caused by its dependence
 on metallicity. As pointed out by \citet{Miller2018} this is a more
 general problem of current simulations.

\begin{figure}[ht!]
 \begin{center}
   \includegraphics[width=0.45\textwidth]{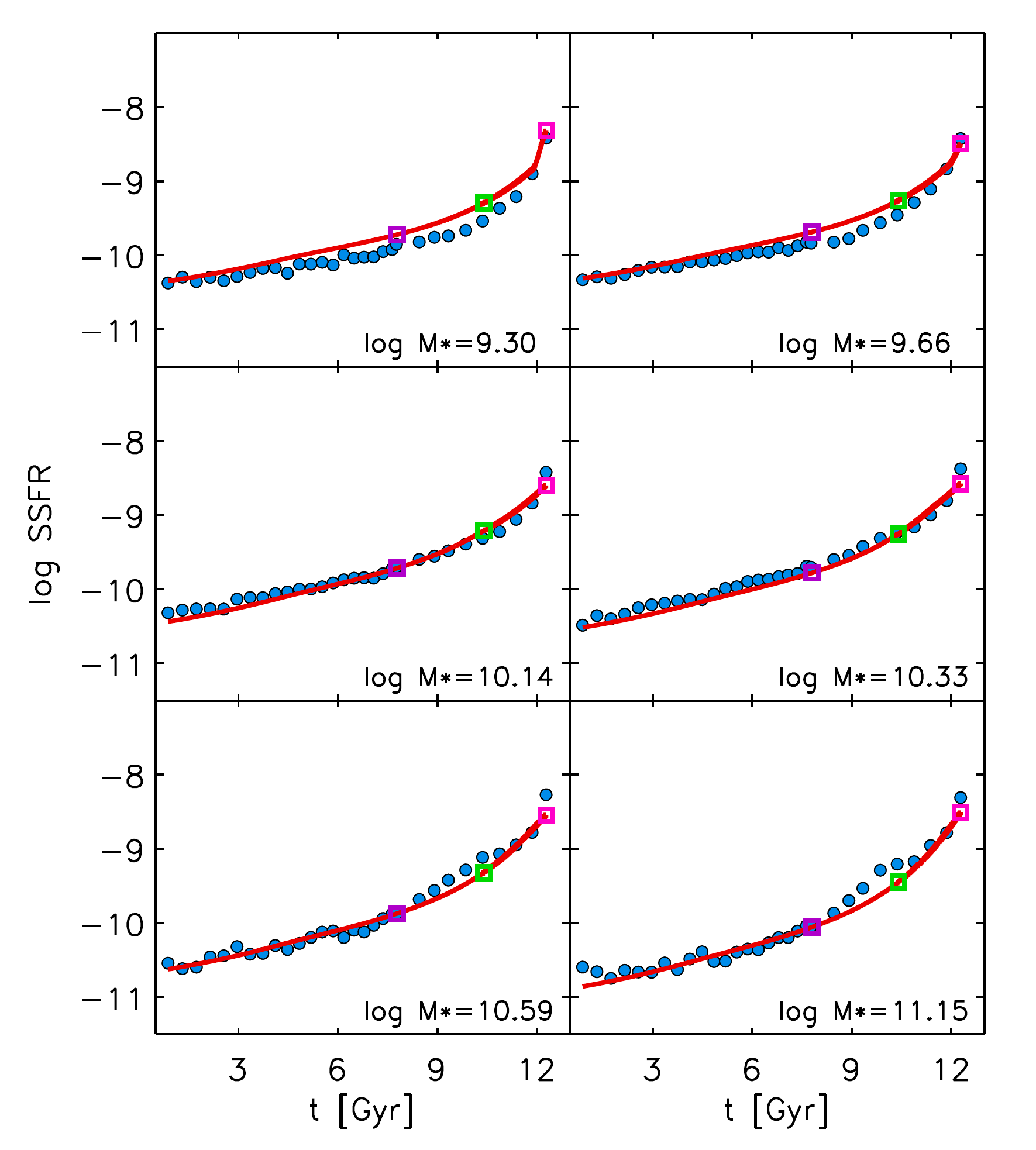}
   \caption{
Specific star formation rate as a function of lookback time for
six galaxies with different final stellar mass (at t = 0). The red
curves correspond to lookback model calculations. The
violet, green and pink squares indicate redshifts z = 1, 2 and 4.2,
respectively. The blue circles correspond to the Magneticum
evolution sample described in the text.
  \label{Fig7} }
 \end{center}
\end{figure}

 We can also compare the time
evolution of the Magneticum evolution sample with the $\psi_{LB}$
of our lookback models. This is done in Fig.~\ref{Fig7}, where
specific SFRs (sSFR) $\psi$/M$_*$ are plotted for models corresponding to
different Magneticum evolution mass bins. We conclude that the Magneticum SFRs are well represented by our choice
 of $\psi_{LB}$.

 We note that in the complete snap shot set of Magneticum simulations there is a significant
 number of disk and intermediate galaxies with very low SFRs. Since
 this work focusses on star forming galaxies not too far from the main
 sequences, we have
 excluded those objects in Fig.~\ref{Fig3} and from all further
 comparison between lookback models and Magneticum. As a selection
 criterion we use the threshold of 0.8 dex below our adopted star
 formation law and include only galaxies above this threshold. We
 realize that such galaxies with low SFRs below our threshold may exist in the real
 universe, but would be excluded from samples selected on strength of
 ISM emission lines.

\section{Comparison between lookback models and Magneticum simulations}

\subsection{The relationship between gas mass and stellar mass}

\begin{figure}[ht!]
 \begin{center}
   \includegraphics[width=0.45\textwidth]{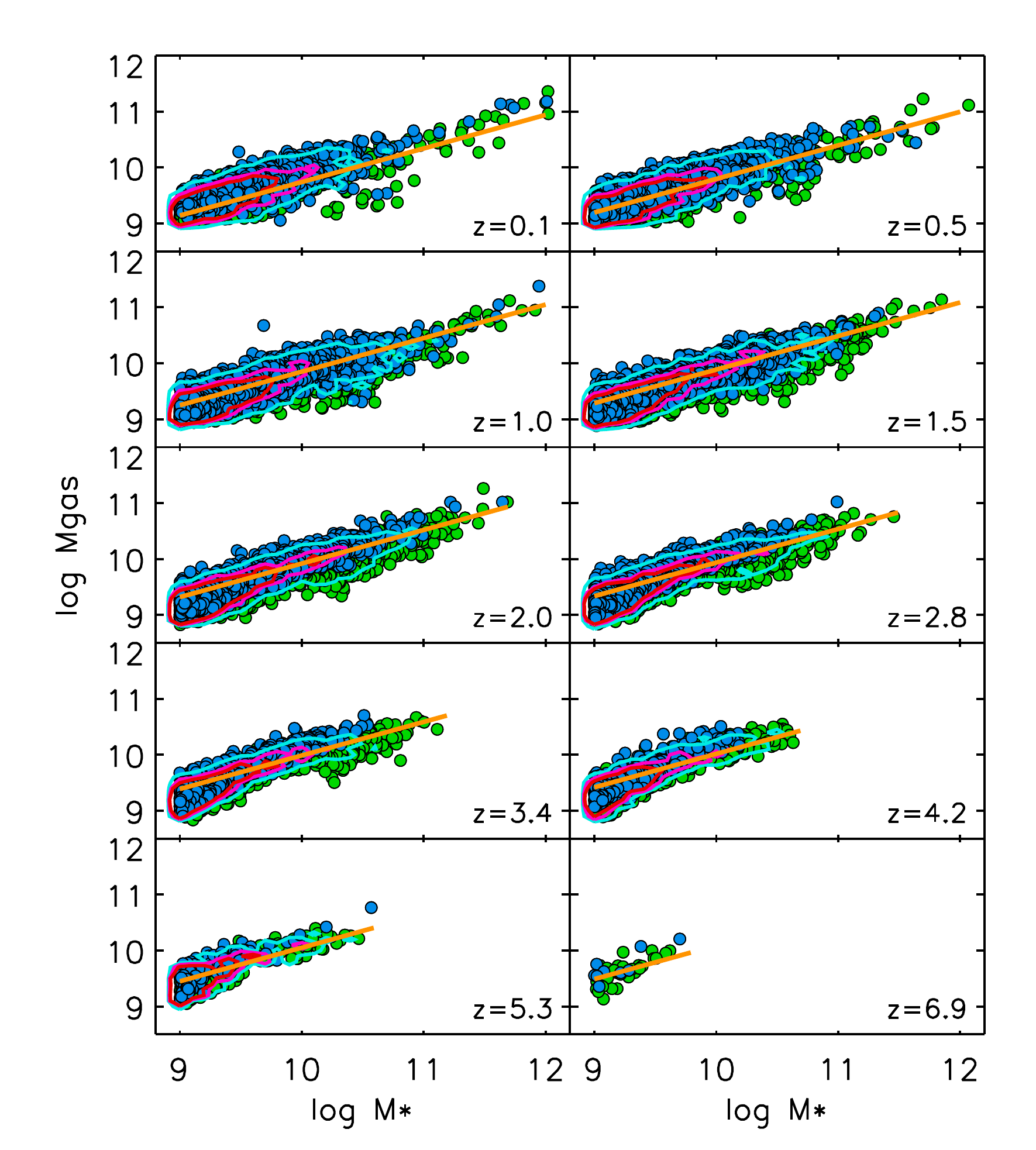}
 \caption{
Logarithm of galaxy cold ISM gas mass versus
logarithm of stellar mass at different redshifts. The Magneticum
galaxies are shown as blue and green circles as before. The
lookback model relationship is plotted as the orange line. As
  in Figure 5 we also show
  isocontours enclosing 68\% (red), 80\% (pink) and 95\% (cyan) of the
  Magneticum galaxies.
  \label{Fig8} }
\end{center}
\end{figure}

The key simplification of the lookback models is the assumption of a redshift dependent
power law between galaxy stellar mass and the total mass of the cold
ISM gas (see eq. 5 and 6). In Fig.~\ref{Fig8}
we test whether the Magneticum hydrodynamic cosmological simulations
support the idea of such a relationship. Using the parameters $\alpha,
\beta$ and log A$_0$ of section 1 we find good agreement with
the Magneticum simulations. This result is very reassuring for the
lookback model approach. It is a strong confirmation of the basic
concept of these models. It also provides a substantial simplification
for the description of star forming galaxy evolution.

We note that the self-consistent analytical approach by \citet{Pantoni2019} and \citet{Lapi2020} discussed in section 2 also leads to a power law relationship between gas mass and stellar mass similar to our lookback models (see Figures 7 in both papers). The redshift dependence found by \citet{Pantoni2019} is in agreement with our models and with Magneticum but it is significantly steeper in the results displayed by \citet{Lapi2020}.

One might argue that for galaxies on the main squence, where star
formation follows a power law with stellar mass, it is to be expected
that ISM gas mass depends on stellar mass in a similar way, because
star formation should be proportional to gas mass. However, the
situation is more complicated, because not all the ISM
gas is involved in the star formation process and the fraction of star
forming to total ISM gas changes
with stellar mass and redshift. We will discuss this further below. 

Nonetheless, we note that there is a weak star formation
dependence of the relationship between ISM gas and stellar mass in the
Magneticum snapshot sample. Galaxies with lower gas mass tend to have
lower SFRs and higher gas mass galaxies have higher SFRs. This has
consequences for the MZR and its SFR dependence, which will be
discussed this further below.

\subsection{Luminosity weighted ages of the stellar population}

\begin{figure}[ht!]
 \begin{center}
   \includegraphics[width=0.45\textwidth]{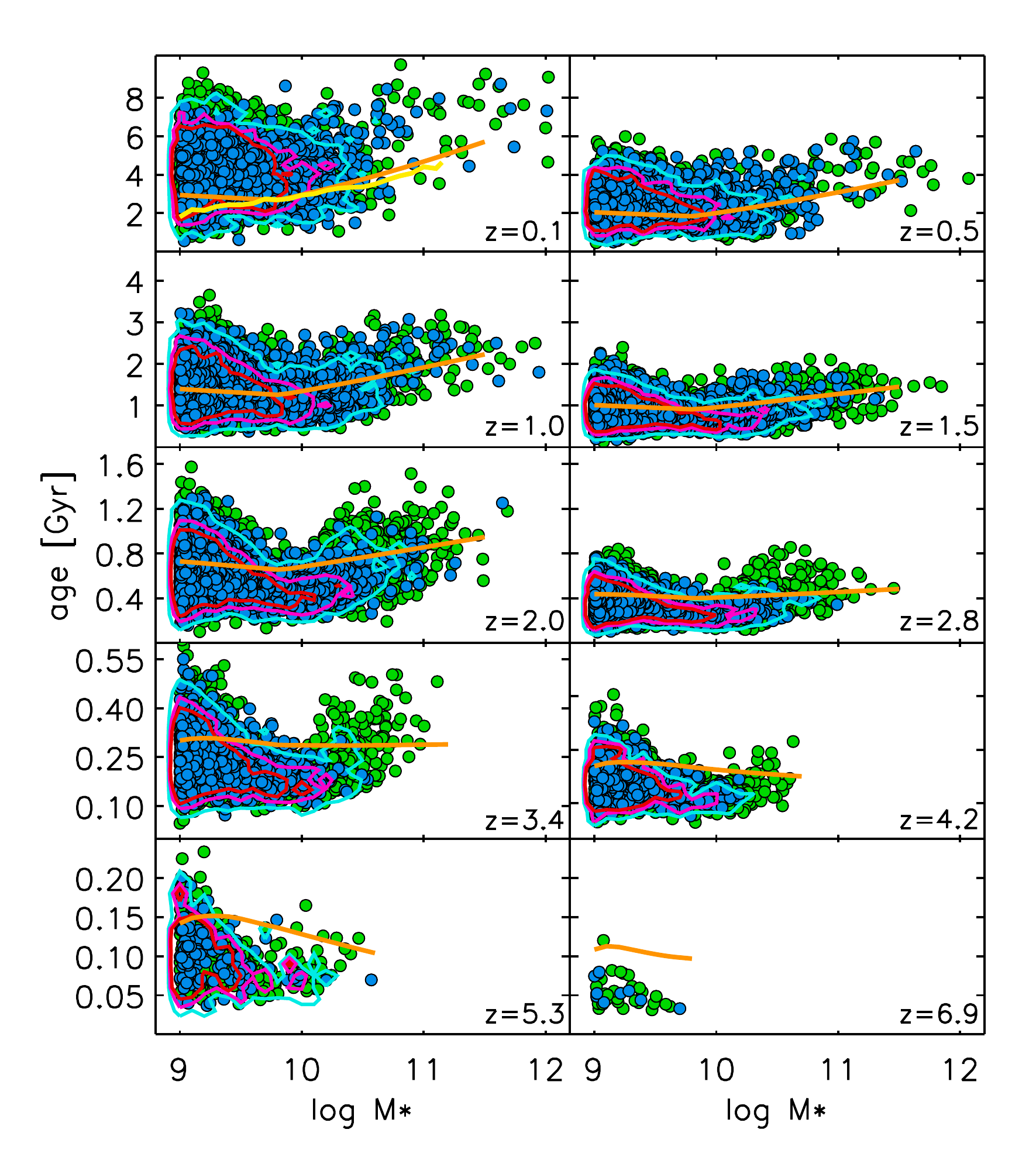}
 \caption{
V-band luminosity weighted average ages (in Gyr) of the Magneticum galaxies
stellar populations (green and blue circles as before) compared with ages calculated from the lookback
models using the standard lookback model SFR law
$\psi_{LB}$ (orange) described in section 5. For the lowest redshift
we also show observational results (yellow) obtained by
\citet{Trussler2020} (see text). Isocontours have the same meaning is
in the previous figures. 
  \label{Fig9} }
 \end{center}
\end{figure}

As the result of their different star formation histories the
distribution of ages of the stellar populations in different galaxies
will vary. A crucial quantity to characterize the age distribution
is the luminosity weighted average age. In Fig.~\ref{Fig9} we show the V-band
luminosity weighted ages of the Magneticum galaxies as a function of
stellar mass for different redshifts and compare with the lookback
model predictions (for the calculation of lookback model luminosity
weighted ages see Appendix C). Except for the very lowest and highest
redshifts the lookback models
seem to agree with the Magneticum simulations. However, we note a
large spread of stellar ages for the Magneticum galaxies. At the
lowest redshift we also compare with observations obtained from a
spectroscopic study of SDSS galaxies by \citet{Trussler2020}, who used
stellar population modelling with FIREFLY to constrain ages of star
forming galaxies (see their section 6.1 and Figure 13). The agreement
with our lookback models is excellent.

\begin{figure}[ht!]
 \begin{center}
   \includegraphics[width=0.45\textwidth]{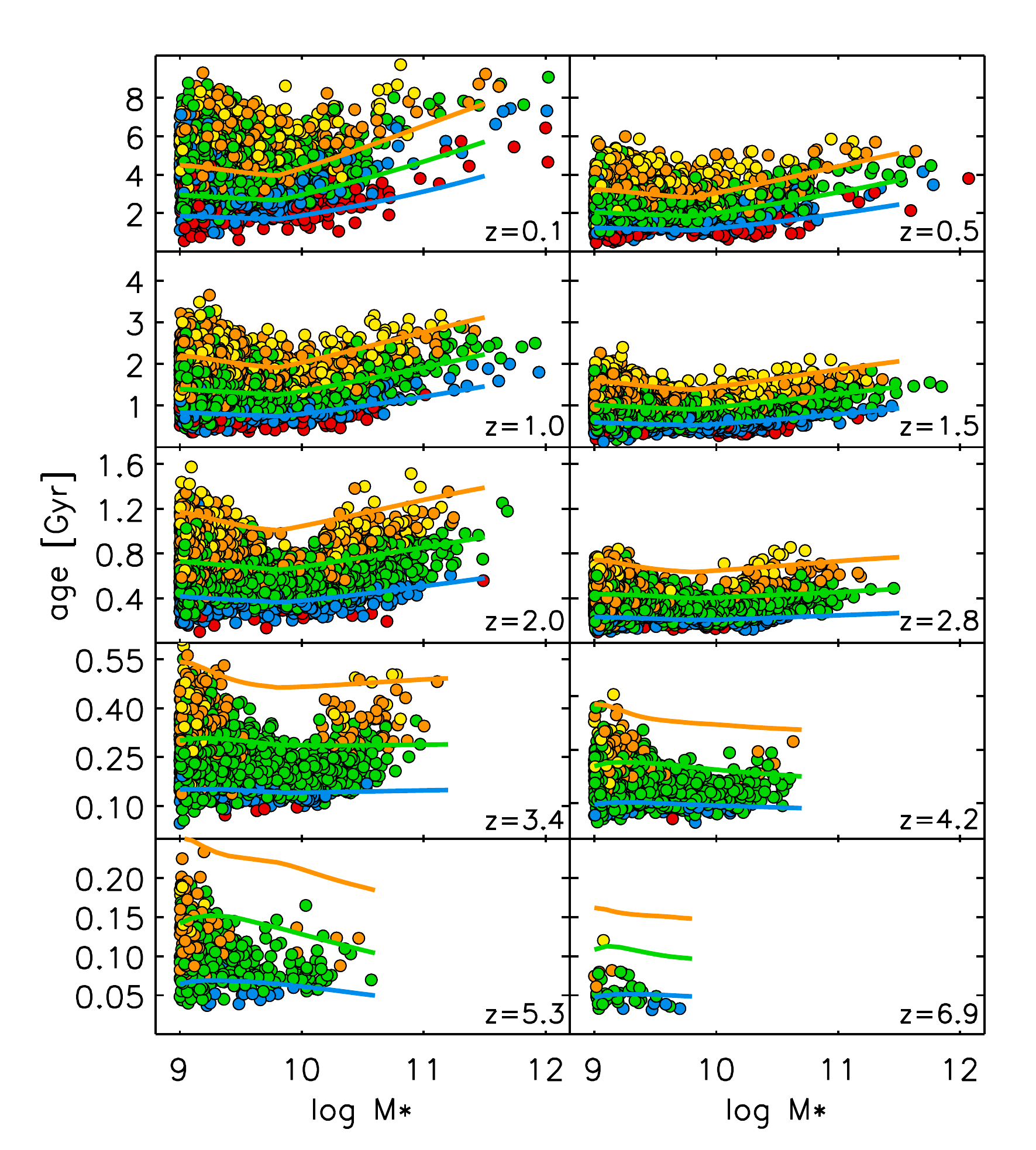}
   \caption{
Same as Fig.~\ref{Fig9} but now the Magneticum galaxies are color-coded
with respect to SFR. Bin1: yellow, bin2: orange, bin3: green, bin4:
blue, bin5: red. Bin 1 corresponds to the lowest SFRs and bin 5 to the
highest (for the extact definition see text). The green curve
corresponds to ages calculated with the standard lookback model SFR law $\psi_{LB}$. The
orange and blue curves are obtained from lookback model calculations,
where log $\psi_{LB}$ is modified by offsets of -0.375 dex and +0.375 dex, respectively.
  \label{Fig10} }
 \end{center}
\end{figure}

Since ages are related to star formation history, it is a crucial
first test to investigate whether the age spread is correlated with
SFRs. From Fig.~\ref{Fig5} we know that our Magneticum snap shot
sample has a wide range of SFRs ($\approx \pm$ 0.8 dex) at fixed
stellar mass. We define five SFR bins for the Magneticum SFRs
$\psi_{Magn}$ with respect to the adopted
lookback SFRs $\psi_{LB}$ in the following way: bin1: log
$\psi_{Magn}$/$\psi_{LB}$ $\le$ -0.5;
bin2: -0.5 $\le$ $\psi_{Magn}$/$\psi_{LB}$ $\le$ -0.25; bin 3:
-0.25 $\le$ $\psi_{Magn}$/$\psi_{LB}$ $\le$ +0.25; bin4: +0.25 $\le$
$\psi_{Magn}$/$\psi_{LB}$ $\le$ +0.50; bin5: +0.5 $\le$
$\psi_{Magn}$/$\psi_{LB}$.

Fig.~\ref{Fig10} shows the ages of the Magneticum galaxies again
but now color coded with respect to SFR bin. There is an obvious
anti-correlation between age and star formation rate, which is easy to
understand. The oldest galaxies at a given stellar mass are mostly
those with the lowest SFRs, because it took a long time to build up the
stellar mass, and the youngest are mostly those with high SFRs,
because the stellar mass was built up recently.

To simulate the effect of systematically higher and lower star
formation rates we have also calculated lookback models with SFRs of log
$\psi_{LB} \pm$ 0.375. The luminosity weighted ages of the stellar
population of these models are also shown in Fig.~\ref{Fig10}. We see
that by accounting for systematically higher or lower SFRs in the
lookback models we obtain ages in agreement with Magneticum.

\subsection{Gas accretion}

In the previous subsections we compared the lookback models with the
Magneticum snap shot sample, the set of simulated star forming galaxies
at different redshifts. Now we turn to the Magneticum evolution
sample, where we combine galaxies in stellar mass bins and then follow
their evolution with time calculating averaged properties of each mass
bin as a function of time (or redshift). In a first step we study the
evolution of galaxy gas mass and investigate the role of gas
accretion. Both the lookback models and the Magneticum simulations include the
effect of gas accretion from the galactic halo or the intergalactic medium.  In the
case of the lookback models this is done implicitly through the
assumption of the power law relationship between gas and stellar mass
M$_g$ = A(z)M$_*^{\beta}$. In the case of the Magneticum models this
is a direct result of the hydrodynamic simulation of the galaxy
formation and evolution process. The change of gas mass with redshift
(or time) is given by

\begin{equation}
\Delta M_g = \Delta M_{\mathrm{accr}}^{\mathrm{eff}} - \Delta M_*
  \end{equation}

 where  $\Delta M_*$ is the change of stellar mass through star formation, which causes a decrease of gas mass.
 $\Delta M_{\mathrm{accr}}^{\mathrm{eff}}$ is the net amount of
 accreted gas mass leading to
 an increase of ISM gas mass and consists of two terms

 \begin{equation}
   \Delta M_{\mathrm{accr}}^{\mathrm{eff}} = \Delta M_{\mathrm{accr}} - \Delta M_{\mathrm{wind}},
 \end{equation}  

where the first describes mass gain through accretion and the second
term mass loss through galactic winds. The normalized effective
accretion rate $\Lambda_{\mathrm{eff}}$ is then

\begin{equation}
 \Lambda_{\mathrm{eff}} = { \Delta M_{\mathrm{accr}}^{\mathrm{eff}}  \over \Delta M_*} = {\Delta M_g \over \Delta M_*} + 1.
\end{equation}

We note that the normalized effective accretion rate is equal to the
effective mass accretion factor, which is frequently used  in chemical
evolution models (see, for instance, \citealt{Kudritzki2015})

\begin{equation}
\Lambda_{\mathrm{eff}} = {\dot{M}_{\mathrm{accr}} \over (1 - R)\psi} -
{\dot{M}_{\mathrm{wind}}  \over (1 - R)\psi}, 
\end{equation}

\hfill \break
where $\dot{M}_{\mathrm{wind}}$ and $\dot{M}_{\mathrm{accr}}$ are the
rates of mass-loss through winds and mass-gain through accretion,
respectively. In the case of the lookback models the normalized
effective accretion rate can be calculated analytically (see section 2)

\begin{equation}
\Lambda_{\mathrm{eff}}  = 1 + \beta{M_g \over M_*}(1 - K(M_*,z)).
\end{equation}

\begin{figure}[ht!]
 \begin{center}
   \includegraphics[width=0.45\textwidth]{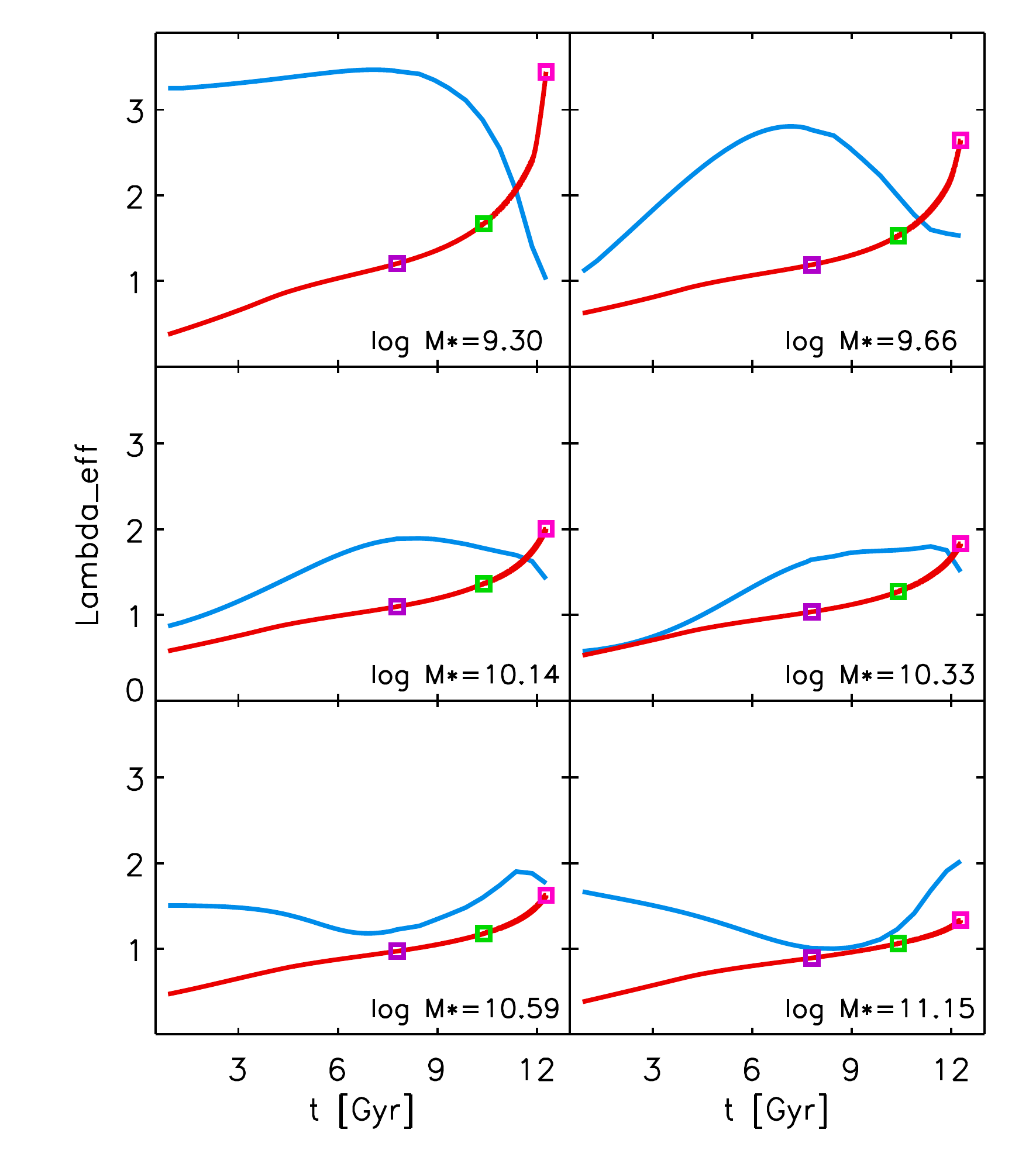}
   \caption{
Normalized effective accretion rate as a function of lookback time for
six galaxies with different final stellar mass (at t = 0). The red
curves correspond to lookback model calculations. The
violet, green and pink squares indicate redshifts z = 1, 2 and 4.2,
respectively. The blue curves are calculated from the Magneticum
evolution sample described in the text.
  \label{Fig11} }
 \end{center}
\end{figure}

Fig.~\ref{Fig11} shows the accretion rates of the lookback models (red
curves) as a function of time for the galaxies with six final masses
log M$_*$. We see that accretion
dominates over galactic winds ($\Lambda_{\mathrm{eff}} >$ 0) but the
mass gains through accretion rates are moderate and do not exceed the SFRs by a
large factor. For the lookback models the value of
$\Lambda_{\mathrm{eff}} \approx$ 1 implies accretion rates half of the
SFRs (note that 1 - R = 0.55 for the look back models). The $\Lambda_{\mathrm{eff}} $
rates of the Magneticum simulations are of the same  order for the four
higher galaxy masses, but are larger by a factor two to three for log
M$_*$ = 9.30 and 9.66, respectively. 

\begin{figure}[ht!]
 \begin{center}
   \includegraphics[width=0.45\textwidth]{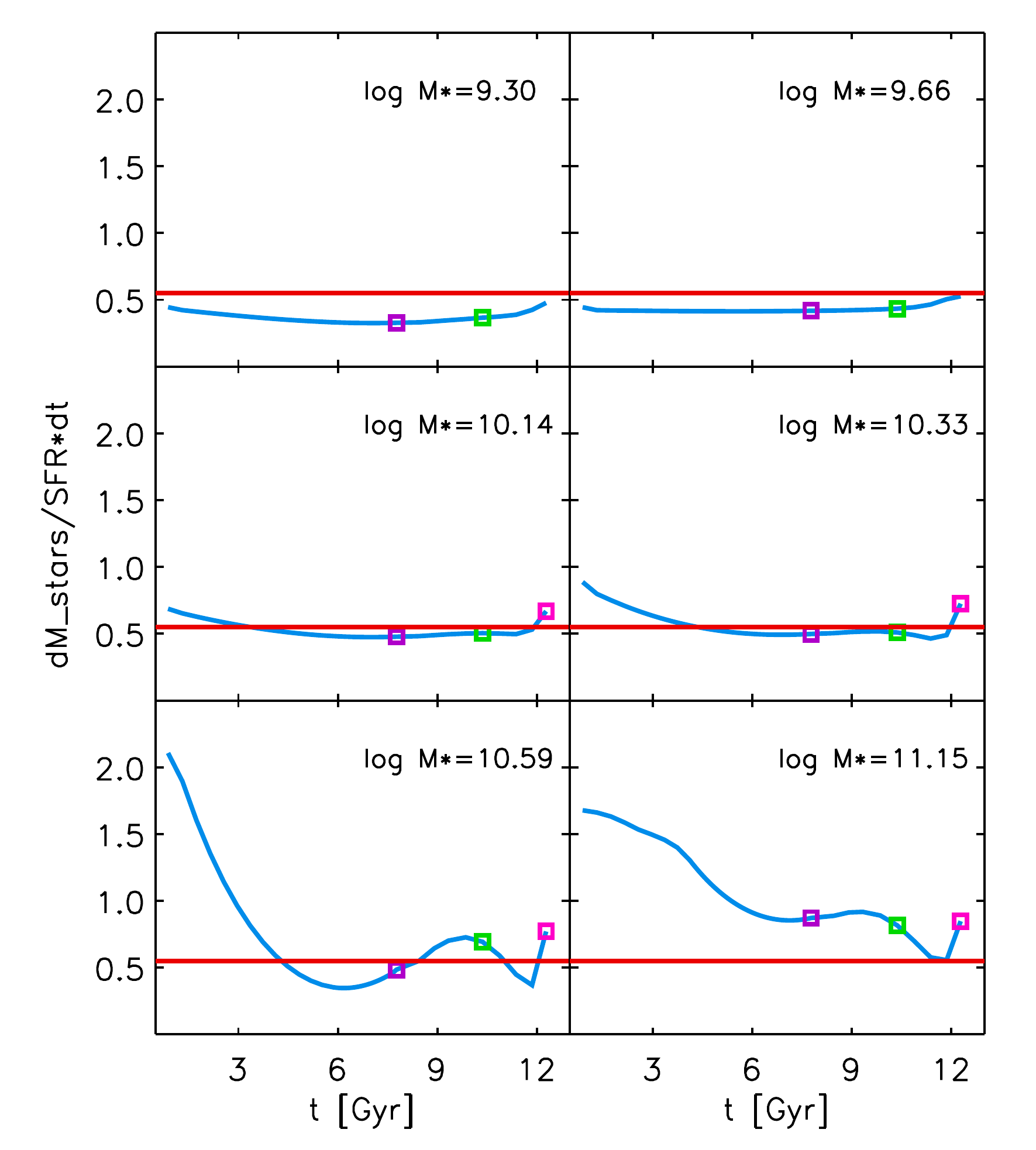}
   \caption{
Changes of stellar mass in units of star formation  as a function of lookback time for
six galaxies with different final stellar mass. The red
curves correspond to lookback model calculations. The blue curves are calculated from the Magneticum
evolution sample. Violet, green and pink squares indicate redshifts z = 1, 2 and 4.2,
respectively. For discussion, see text.
  \label{Fig12} }
 \end{center}
\end{figure}

While the differences between the lookback models with the simple
assumption of a redshift dependent power law relationship between gas
mass and stellar mass and the detailed Magneticum hydrodynamic
simulations are obvious, we note that they are not orders of
magnitude. We take this as an additional confirmation of the lookback
model approach. As we will see in the next subsection, this is mostly
caused by somewhat higher values of R and not so much by higher
accretion rates.

However, at this point we need to add a word of caution. We note that our method to calculate Magneticum accretion
rates is based on the assumption that the increase of stellar mass is
solely through star formation. If the increase of stellar mass is also
partially caused by merging with infalling galaxies, then our approach
underestimates the accretion rates and provides only lower limits.

\subsection{Stellar mass growth and merging}

In order to estimate the influence of merging on the growth of stellar
mass we determine the ratio

\begin{equation}
\Lambda_* = {\Delta M_* \over \psi \Delta t}.
\end{equation}

For the lookback models we have the constant value $\Lambda_*$ = 1 -
R. This is the horizontal red line in Fig.~\ref{Fig12}. The blue curve
shows the results obtained from the Magneticum evolution sample. We
find good agreement with the simple lookback model approach for the
mass bins log M = 10.16 and 10.34. This implies mass return fractions  to the ISM during the star formation
process of R $\approx$ 0.45, as we have adopted for the lookback
models. For smaller masses the blue curves for the Magneticum models
indicate somewhat larger return fractions of R $\approx$ 0.6 to 0.7. For the two highest mass bins there is a clear indication
of merging at lookback times t $\lesssim $ 3 and 6, respectively, with
$\Lambda_*$-values clearly larger than unity. As a consequence, the
Magneticum accretion rates for these phases of the evolution of the
Magneticum high mass models are likely larger than indicated in Fig.~\ref{Fig11}.

\subsection{Magneticum star forming gas masses, star formation time
  and the main sequence star formation law}

The Magneticum simulations provide the opportunity to investigate the
connection between the main sequence star formation law $\psi$(z,M$_*$)
and the gas mass - stellar mass relationship M$_g$(z,M$_*$). We do
this in three steps. We first discuss the fraction of cold ISM
gas, which is contributing to star formation. Second, we look at the time
scale of the star formation process and, third, we combine this
information with the predicted relation of total cold ISM gas mass
(star forming and passive with respect to the star formation process) with stellar mass.

\begin{figure}[ht!]
  \begin{center}
    \includegraphics[width=0.45\textwidth]{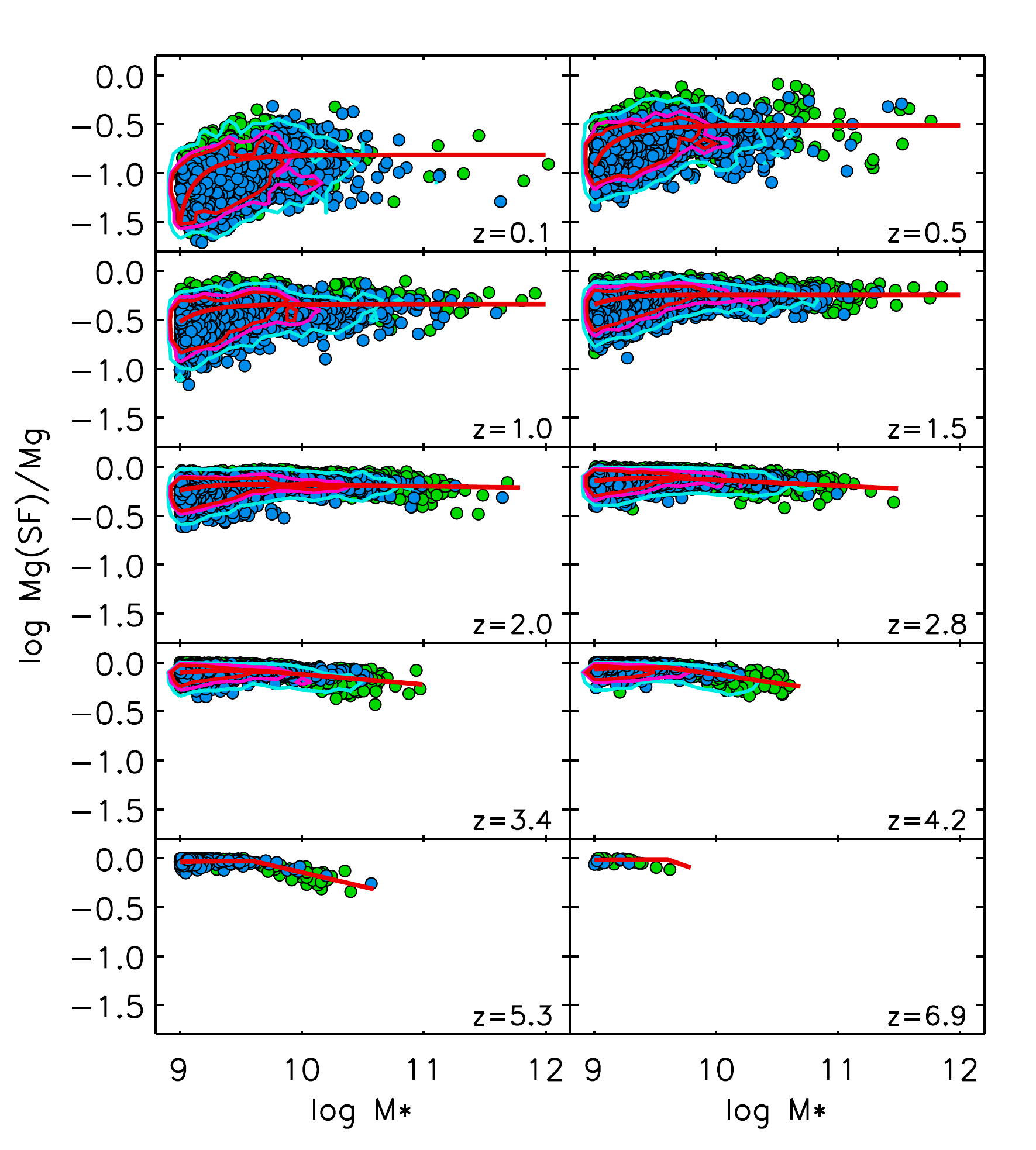}
   \includegraphics[width=0.45\textwidth]{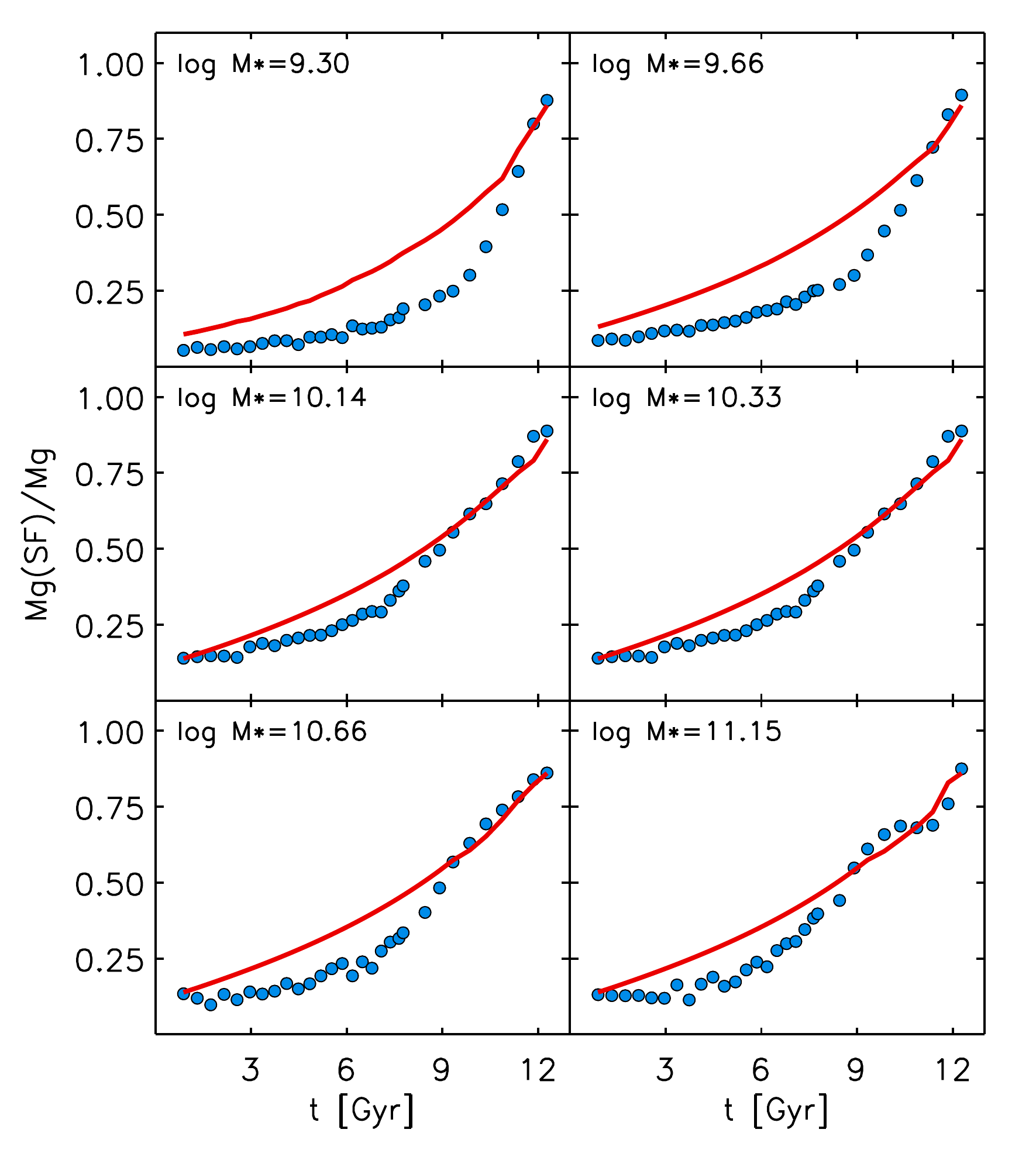}
   \caption{
ISM fraction of the mass of star forming gas to total gas
mass. Top: Magneticum snap shot sample at different redshifts with log
M$_g$(SF)/M$_g$ as a function of stellar mass. Blue and green circles and isocontours
as before. Bottom:  M$_g$(SF)/M$_g$ as a function of lookback time for
the Magneticum evolution sample (blue circles) in six different stellar mass bins. The red
curves correspond to the fit described in the text. A formula is
given in Appendix B.
  \label{Fig13} }
 \end{center}
\end{figure}

Not all the cold ISM gas in a galaxy is involved in the star forming
process. In the Magneticum simulations only a fraction x$_g$ 

\begin{equation}
x_g(z(t),M_*) = {M_g^{SF}(z,M_*) \over M_g(z,M_*)}
\end{equation}

 is producing stars. M$_g^{SF}$ is the total mass of all cold star forming ISM
 gas and correponds to gas of low temperature above a certain density
 threshold. Fig.~\ref{Fig13} uses the Magneticum snap shot sample
 (restricted to galaxies with log $\left|\psi/\psi_{LB}\right| \leq$ 0.25, the
 Magneticum main sequence) and the evolution sample to
 demonstrate how x$_g$ depends on redshift and stellar mass and how it
 evolves with lookback time t.

 The snap shot sample plots show a significant scatter at lower
 redshift. However, observations of galaxies in the local universe
 show a similar scatter and the x$_g$ values observed agree with the
 Magneticum observations (see the XGASS, xCOLDGASS and MAGMA surveys,
 \citealt{Saintonge2017}, \citealt{Catinella2018}, \citealt{Hunt2020}).

\begin{figure}[ht!]
  \begin{center}
    \includegraphics[width=0.45\textwidth]{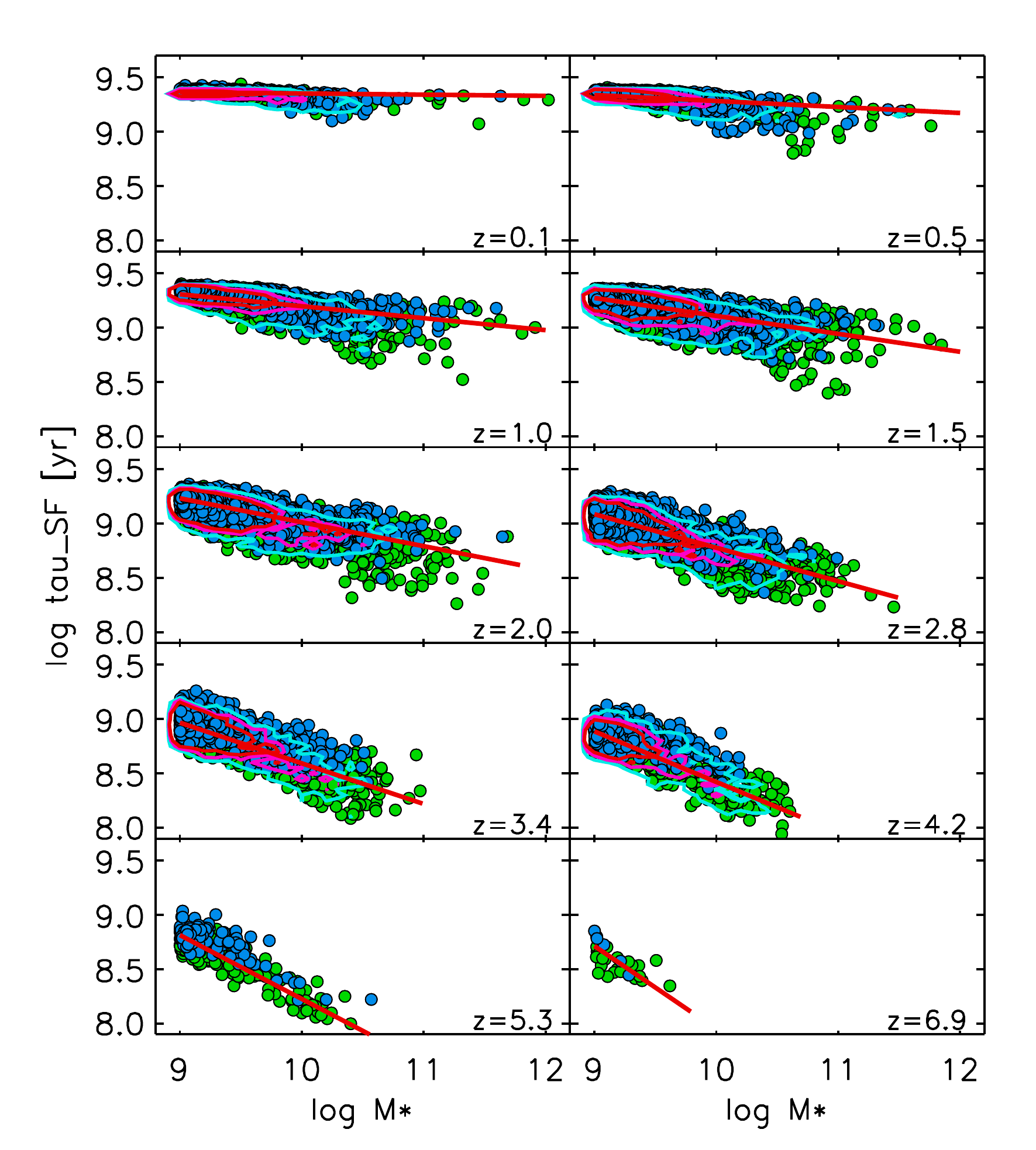}
    \includegraphics[width=0.45\textwidth]{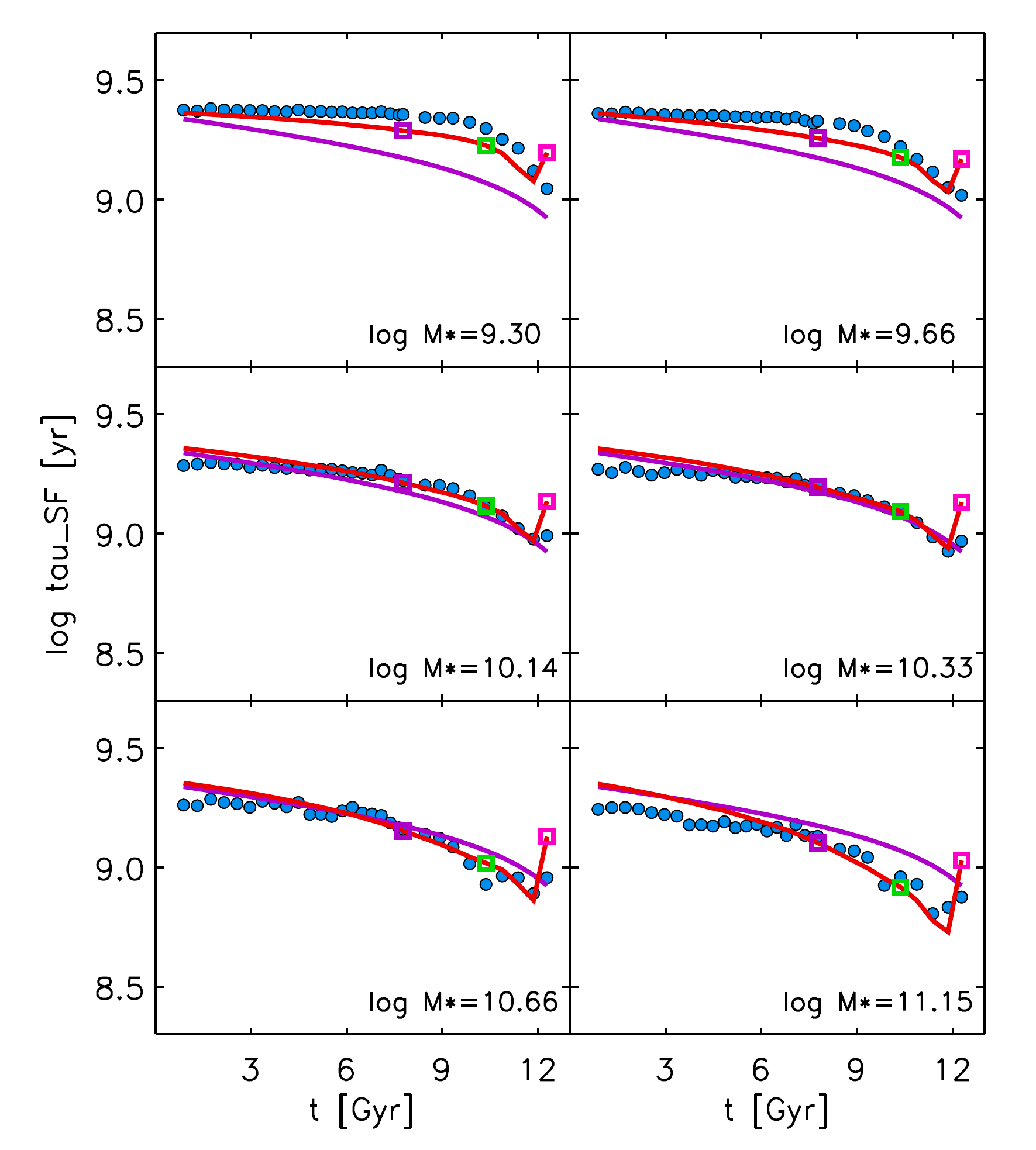}
   \caption{
Logarithm of star formation time. Top:  Magneticum snap shot sample at
different redshifts with log $\tau_{SF}$ as a function of stellar
mass. Blue and green circles \textbf{and isocontours}
as before. Bottom: $\tau_{SF}$  versus lookback time for the
Magneticum evolution sample in six different stellar mass bins (blue
circles). The red curves are calculated according to Appendix B and are
discussed in the text. The violet curves are given by Eq. (26).
  \label{Fig14} }
 \end{center}
\end{figure}
 
 The red curve is calculated
 with a fit formula (see Appendix B) to the Magneticum data, which closely matches the mean
 of x$_g$ as a function of stellar mass at each redshift. At
 low redshift x$_g$  is constant with stellar mass except at the
 lowest masses, where we find a strong increase. Towards higher
 redshift this behavior reverses: x$_g$ is constant at low masses,
 but declines towards higher masses. Most importantly, though, the
 maximum of x$_g$ increases with redshift. Unfortunately, there are no
 direct observations of HI available in galaxies of higher redshift
 to compare with the Magneticum simulations.

 The evolution sample plots  demonstrate very clearly that during the
 course of the evolution of a galaxy x$_g$ decreases continuously.
At the beginning of the life of a Magneticum star forming galaxy
practically all gas is involved in the star formation process. But
then the fraction of gas contributing to star formation becomes
significantly smaller. The red curve is the fit taylored to describe
the snap shot sample. While it is not a perfect fit for the evolution
sample, it captures the evolution with time (and redshift) reasonably
well.

We note that in observational studies - because of the lack of direct
HI observations - the assumption is frequently made for the evolution
of x$_g$ that already at z
$\approx$ 0.4 all cold ISM gas is involved in the star formation
process, i.e. x$g \approx$ 1 (see, for instance, \citealt{Tacconi2018,Tacconi2020}).
The Magneticum simulations do not support this
assumption. While a detailed comparison is beyond the scope of this
paper, we note that other cosmological simulations (\citealt{Lagos2015},
  \citealt{Diemer2019}, \citealt{Dave2020}) agree with Magneticum in this regard.

The timescale of the star formation process is given by

\begin{equation}
\tau_{SF}(z(t),M_*) = {M_g^{SF}(z,M_*) \over \psi(z,M_*)}.
\end{equation}

Fig.~\ref{Fig14} shows $\tau_{SF}$ for the main sequence galaxies of
the  Magneticum snap shot sample and the evolution sample. The red
line in the plots is a simple fit to the ridge line in the snap shot
sample. The corresponding fit formula is given in Appendix B. We
see that in the Magneticum galaxies the star formation time decreases
with redshift. We also find a mass dependence with a negative slope,
which becomes steeper with redshift. The Magneticum
simulations at z $\sim$ 0.1 agree well with observations of galaxies
in the local Universe (xGASS, xCOLDGASS, MAGMA). Observations at
larger redshift \citep{Tacconi2018,Tacconi2020} are also in agreement
on average but they do not show the pronounced negative slope with
stellar mass.

Comparing with a large sample of observed star forming galaxies
\citet{Tacconi2018, Tacconi2020} suggest an empirical relationship of
the form

\begin{equation}
\tau(t) = \tau_0(1 + z(t))^{-n},
\end{equation}

which we show in the plot of the Magneticum evolution sample in
Fig.~\ref{Fig14} using $\tau_0$ = 2.3$\times$10$^9$ yrs and n=0.6
(violet curve).  We
see that the Magneticum simulations have a similar trend as a
function of lookback time (or redshift). The fit obtained from the
Magneticum snap shot sample is in agreement with eq. (26). We note that
our value for $\tau_0$ is a factor of two larger than the one found by
\citet{Tacconi2018}. This is a
consequence of the fact that the Magneticum SFRs are lower than
the ones used by \citet{Tacconi2018} in the range of lookback time displayed here.

\begin{figure}[ht!]
 \begin{center}
   \includegraphics[width=0.45\textwidth]{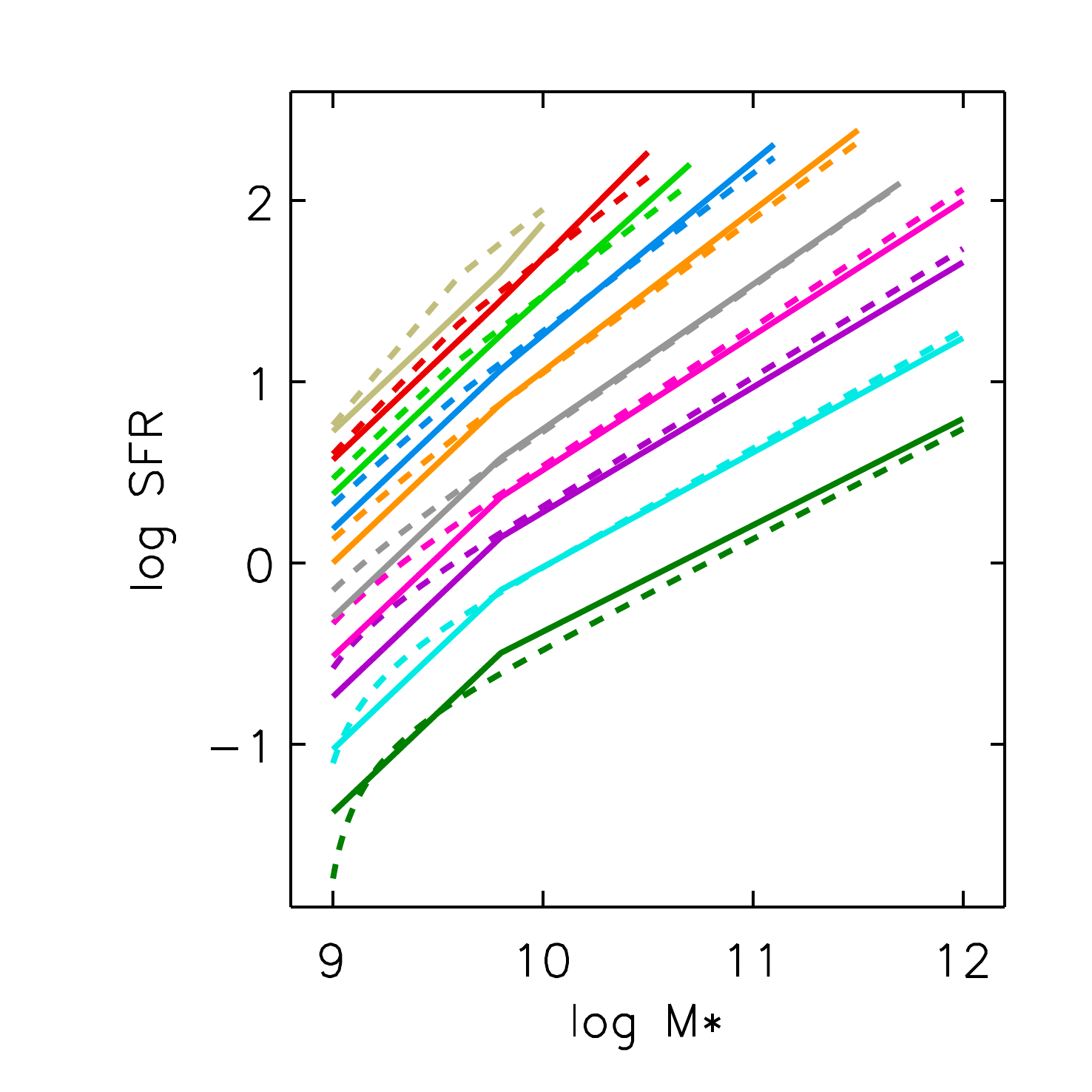}
   \caption{
Lookback model SFRs $\psi_{LB}$ as a function of stellar mass at
the ten redshifts of the Magneticum snap shot sample from z = 0.1 to
6.9 (solid curves) together with SFRs calculated from eq. (27) (dashed).
  \label{Fig15} }
 \end{center}
\end{figure}

With the fits for x$_g$ and $\tau_{SF}$ as given in Appendix B we have
an alternative way to calculate SFR as a function of redshift and
stellar mass

\begin{equation}
\psi(z(t), M_*) = {x_g^{fit}(z,M_*) \over \tau_{fit}(z,M_*)}M_g(z,M_*),
  \end{equation}

where M$_g$(z,M$_*$) is given by eq. (5). Fig.~\ref{Fig15} compares lookback SFRs $\psi_{LB}$ with SFRs
calculated with eq. (27). We use the overall agreement found in
Fig.~\ref{Fig15} as an argument that the SFR law along the main
sequence is a consequence of the power law relationship between total ISM gas mass and stellar
mass. However, it is important to stress that the complex behavior as a
function of redshift and stellar mass of x$g$, the ratio of star
forming to total gas mass, and of $\tau_{SF}$, the star formation
timescale are also of crucial importance. Most importantly,
we note that the decline of specific star formation rates (see Fig.~\ref{Fig7})
is not caused by the relatively small drop of the ratio M$_g$/M$_*$
but rather by the significant continuous decrease of x$_g$(t). In other
words, galaxies form less stars in the course of their evolution not
because they are running out of gas (or because the star formation
time changes dramatically) but rather because the fraction of cold ISM
gas, which is capable of producing stars, becomes smaller.

\subsection{Chemical evolution}

\begin{figure}[ht!]
 \begin{center}
   \includegraphics[width=0.45\textwidth]{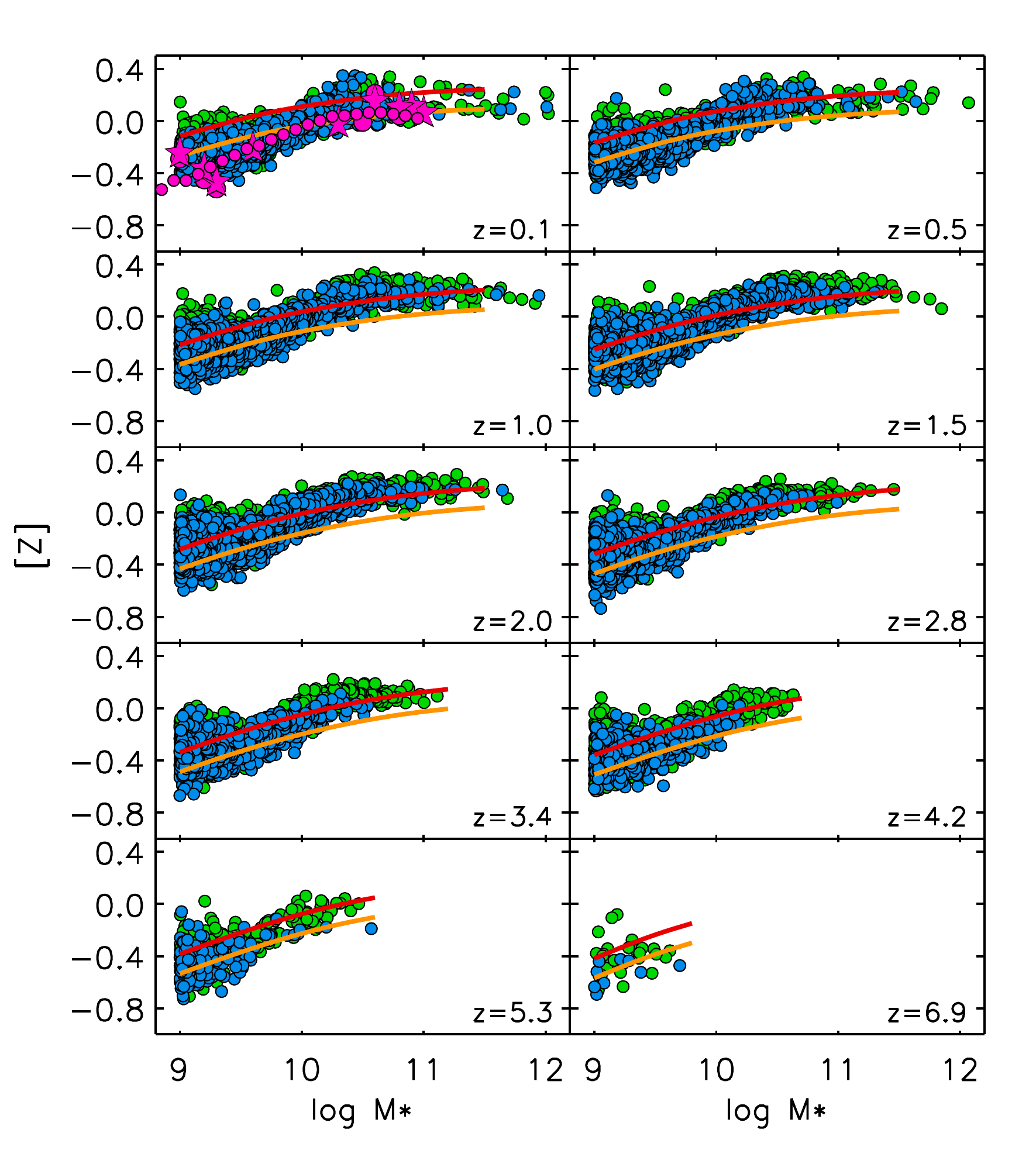}
   \caption{
Luminosity V-band averaged stellar metallicities versus total stellar
mass for the galaxies of the Magneticum snap shot sample at ten
different redshifts. Lookback model stellar metallicities calculated
for two different yields are overplotted in red and orange. At the
lowest redshift observed stellar metallicities (see Fig.~\ref{Fig2})
are also shown as small and large yellow circles and
asterisks. Magneticum isocontours corresponding to the same
  values as in the previous figures are also shown.
  \label{Fig16} }
 \end{center}
\end{figure}

One of the major goals of our lookback models has been to develop a
tool to understand and interprete the chemical evolution of star forming
galaxies. Here the key observation is the MZR  between stellar
metallicity and total stellar mass.  Fig.~\ref{Fig16} shows the V-band
luminosity weighted average metallicity of the stellar population
of the Magneticum snap shot sample as a function of stellar mass at
different redshifts. The stellar metallicities of our lookback models
calculated for two different effctive yields  [Z]$_0$ = log${Y_N \over
  Z_{\odot}(1-R)}$ = 0.10 (orange) and 0.25 (red) are also shown. There is a
significant scatter of $\sigma_{[Z]} \approx$ 0.1 to 0.15 dex in the
Magneticum sample, but on average the lookback models  reproduce the
Magneticum MZRs reasonably well. Note that $\alpha$ = 0.4, $\beta$ =
0.6 and log A$_0$ = 3.73 are used for the comparison with the
Magneticum simulations. 

 \begin{figure}[ht!]
 \begin{center}
   \includegraphics[width=0.45\textwidth]{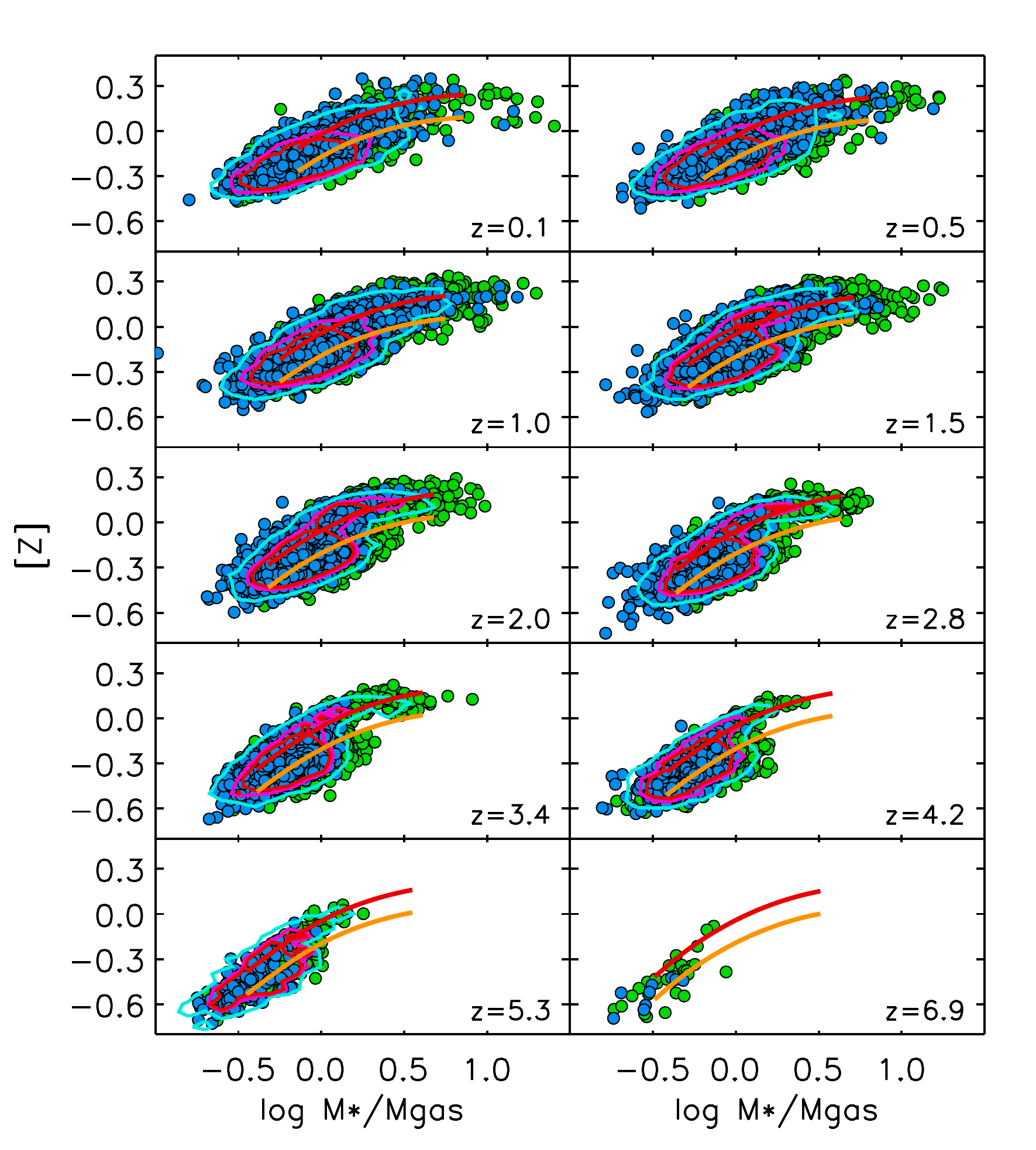}
   \caption{
Stellar metallicities as a function of the ratio of stellar mass to
total ISM gas mass. The galaxies of the Magneticum snap shot sample
are compared with lookback model stellar metallicities again calculated
for two different yields and overplotted in red and
orange. Isocontours use the same values as in  previous figures.
\label{Fig17} }
 \end{center}
\end{figure}

At the lowest redshift we also plot observed stellar metallicities
which are systematically lower than Magneticum and which were fitted in
section 3 with an effective yield lower by 0.025 and 0.175 dex,
respectively. This systematic difference is slightly larger than
the uncertainty of the observational stellar metallicity zero points which are
about 0.1 dex (see references in section 3), but can be addressed by
changes of the yields in the simulations as demonstrated in
Fig.~\ref{Fig16}.

An important consequence of the simplifying assumptions of our
lookback models is the prediction that galactic metallicity is
basically a function of the ratio of stellar mass to ISM gas
mass. Fig.~\ref{Fig17} confirms this conclusion by comparing lookback
model stellar metallicities with Magneticum metallicities as a
function of log M$_*$/M$_g$.

\begin{figure}[ht!]
 \begin{center}
   \includegraphics[width=0.45\textwidth]{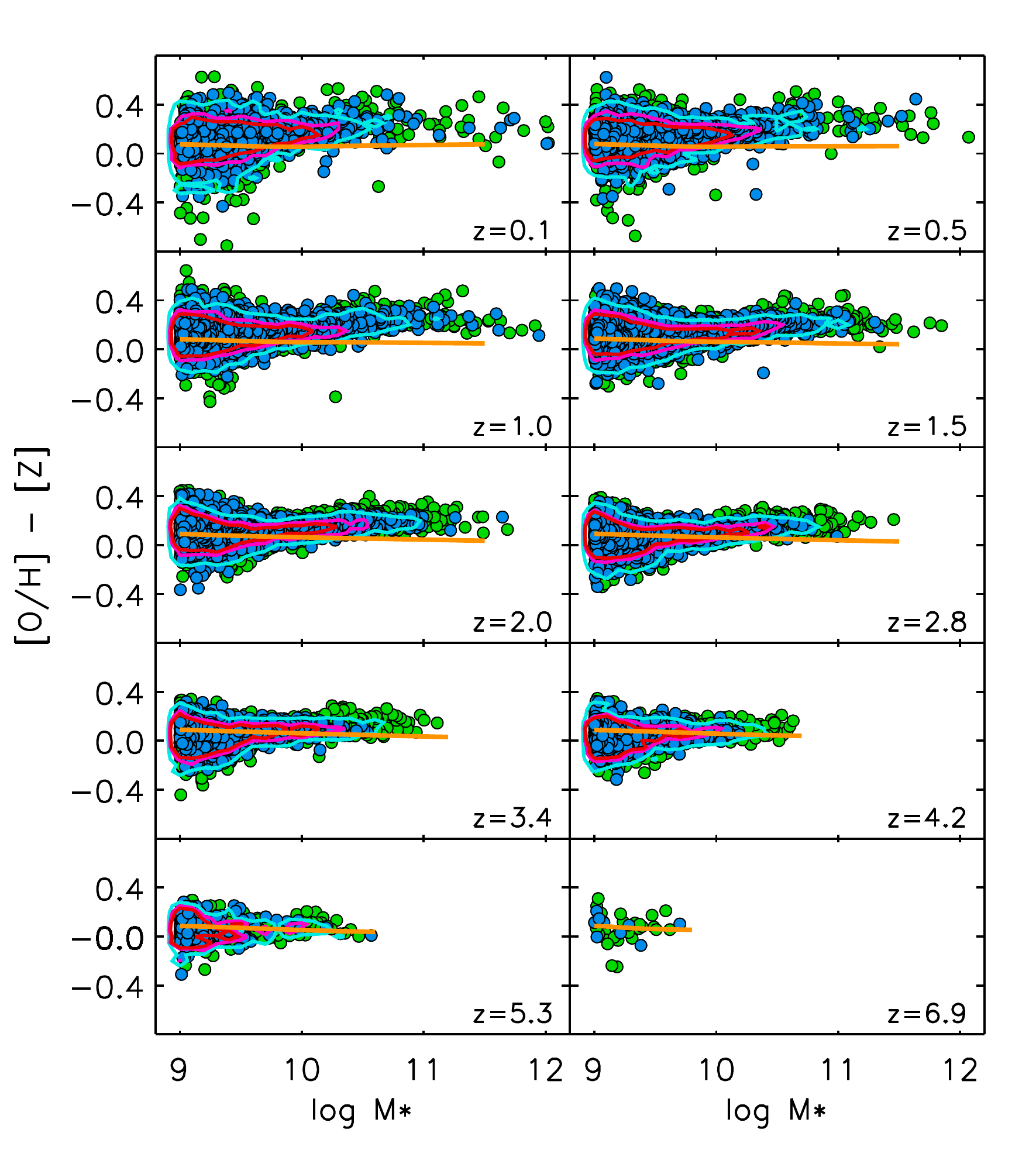}
   \caption{
Difference between the oxygen abundance (relative to the sun) [O/H] of
the star forming ISM gas and the luminosity weighted stellar
metallicity [Z] (also relative to the sun) as a function of galaxy
stellar mass. The galaxies of the Magneticum snap shot sample
are plotted as circles in the usual way and isocontours for the same
values as in previous figures are also shown. The results obtained
from the lookback model stellar metallicities are given by the orange line. 
  \label{Fig18} }
 \end{center}
\end{figure}

The metallicity of the star forming ISM gas is expected to be slightly
higher than the luminosity weighted metallicity of the stellar
population, because the ISM metallicity represents the latest stage of
the chemical evolution, whereas the stellar luminosity weighted
metallicity always contains a contribution by the older population
less advanced in the formation of heavy elements. Fig.~\ref{Fig18} confirms this expectation. Note that we represent the
metallicity of the star forming gas by its oxygen abundance relative
to the sun [O/H]$_{\mathrm{ISM}}$, which is usually determined from the observation of strong emission
lines of the star forming gas. The difference $\Delta$ = [O/H]$_{\mathrm{ISM}}$
- [Z] is somewhat larger than the lookback models for most of the
Magneticum galaxies but the effect is not large (0.1 to 0.15 dex) and
disappears towards higher redshifts. We note that a small fraction of the
Magneticum galaxies has negative values of $\Delta$. We interprete
these as cases of recent accretion or merging with metal poorer gas
involved.

\begin{figure}[ht!]
 \begin{center}
   \includegraphics[width=0.45\textwidth]{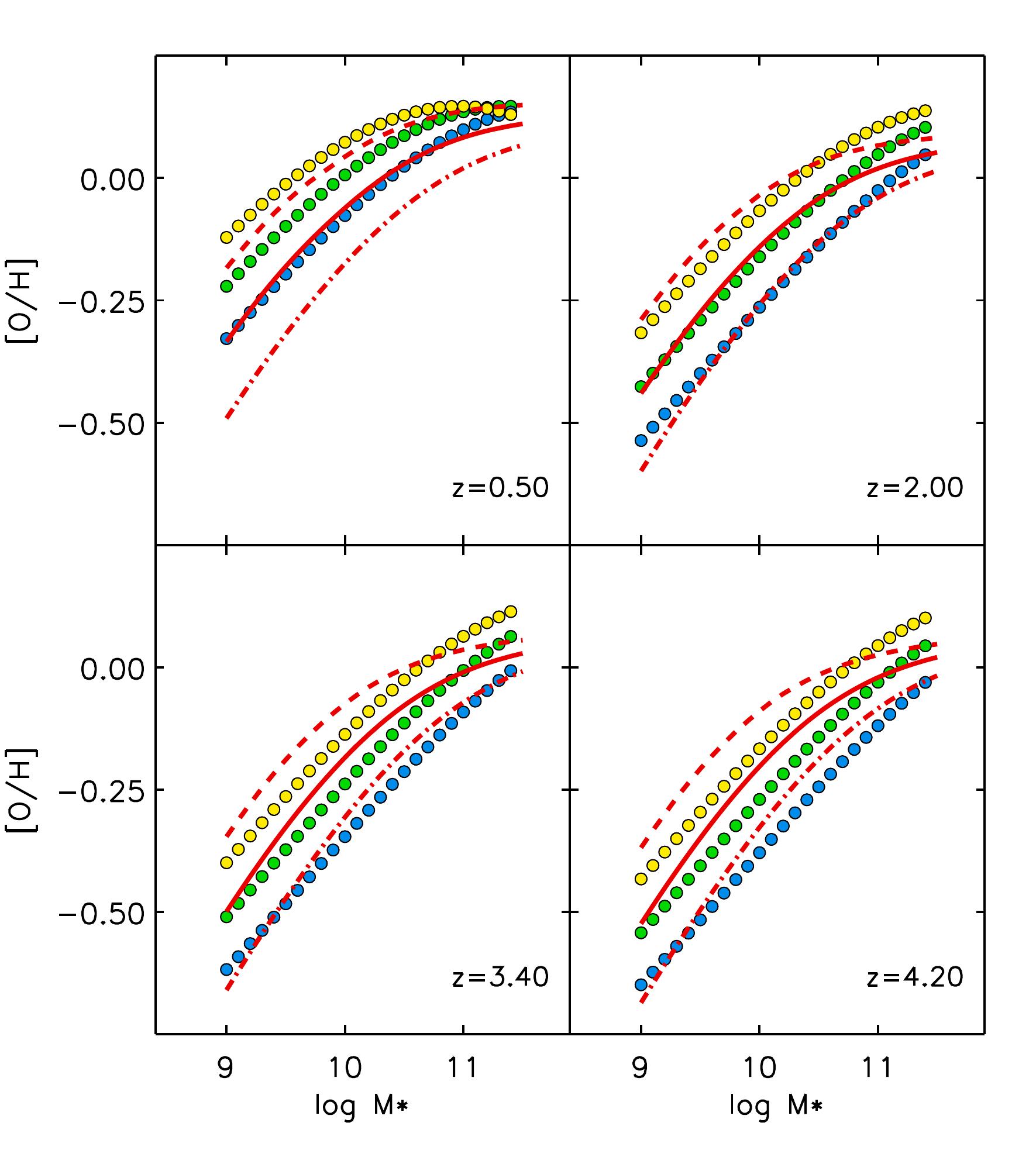}
   \caption{
The observed effect of SFR on the MZRs at fixed redshift. Gas phase
HII region oxygen abundances (relative to the sun) versus
galaxy stellar mass are displayed at four redshifts. The circles
correspond to observations and are obtained from
the \citet{Sanders2020} FMR relationship adopting their SFR main
sequence law (green) and systematic shifts away from the main sequence by $\pm$
0.375 dex (yellow and blue, respectively). The red curves correspond
to lookback model calculations on their main sequence (solid, see eq. 17 and
Appendix B) and with the star formation shifts as described in the
text (dashed and dashed-dotted).
  \label{Fig19} }
 \end{center}
\end{figure}

As was discovered by \citet{Mannucci2010} from a study of HII
region emission lines, the MZR contains a
dependence on a third parameter, the SFR. At fixed stellar mass, galaxies
with lower SFR tend to have higher metallicities and vice versa. Z17  in their investigation of
spectra of the integrated stellar population of 250000 SDSS galaxies
found a similar effect for stellar metallicities. Most recently,
\citet{Sanders2020} investigating galaxy gas-phase
metallicities out to z $\sim$ 3.3, combined their SFRs
and metallicities  and, following the original work by
\citet{Mannucci2010}, presented a new fundamental relationship (FMR)
between metallicity, stellar mass, star formation rate and
redshift. In this framework, the evolution of the MZR with redshift is
a consequence of the influence of SFR, which increases
with redshift and leads to a decrease of metallicity. In addition, at
a fixed redshift the range of SFRs encountered for galaxies at the
same stellar mass contributes to the vertical width of the MZR with
galaxies with lower SFRs having higher metallities and vice versa. We give an example in Fig.~\ref{Fig19}.

In our lookback model approach metallicities do not depend directly on
SFR. As explained in Appendix A, they depend foremost on
M$_*$/M$_g$, the ratio of stellar to ISM gas mass and the
evolution of the MZR with redshift is a result of the fact that the
relation between gas and stellar mass is redshift dependent. Thus at
first glance, the
lookback models seem to be incapable in reproducing the observed SFR
dependence of the MZRs at different redshift.

However, observations of galaxies in the local Universe show that the
power law relationship between gas and stellar mass also depends on
SFR. Galaxies with higher SFR have a higher ratio of gas mass
to stellar mass and vice versa. \citet{Hunt2020} from the study of
their MAGMA sample of galaxies
find that the shift with SFR can be described by
$\Delta$log M$g$ = x$_{\delta}\Delta$log $\psi$ with x$_{\delta}$ =
0.37 (see their eq. 10) .

It is not surprising that such an additional SFR dependence
exists. Galaxies with a higher (lower) gas mass at similar stellar mass and
redshift will very likely also have an increased mass of the star
forming cold ISM gas and, in consequnce, their SFRs will be higher (lower).

For our lookback models this has an important
consequence. We obtain a SFR dependence of the MZR by assuming
that galaxies with a systematic shift away from the lookback model
main sequence by $\delta$ = log $\psi$/$\psi_{LB}$ have also a shift in
the zero point of the gas mass stellar mass relationship
by $\Delta$log A$_0$ = x$_{\delta}\delta$ (see eq. 5 and 6 for the meaning
of A$_0$). This is demonstrated in
Fig.~\ref{Fig19}, where we compare our lookback models with the \citet{Sanders2020} MZRs
calculated from their FMR formula. We adopt $\delta$ = $\pm$0.375 in this
plot and x$_{\delta}$ = $\pm$0.5. In this way, the observed SFR dependence
of the MZRs at fixed redshift is described reasonably well, as
Fig.~\ref{Fig19} indicates.

\begin{figure}[ht!]
 \begin{center}
   \includegraphics[width=0.45\textwidth]{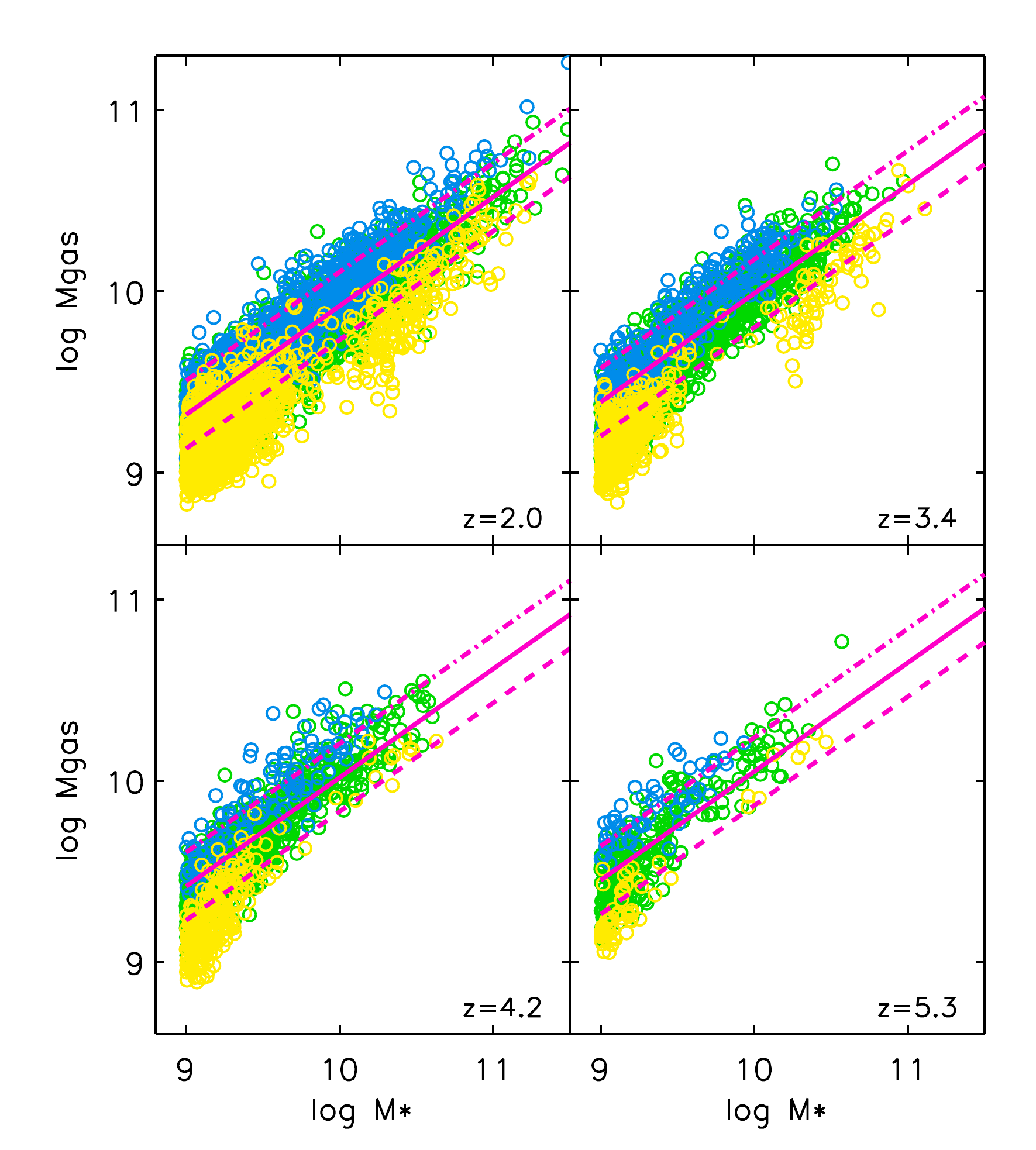}
   \caption{
Magneticum galaxy gas mass versus stellar mass at four redshifts. The
galaxies are color coded according to their SFRs. Green circles
correspond to galaxies around the main sequence within 0.2 dex,
yellow and blue circles are below and above the mains sequence,
respectively. The pink solid line corresponds to the
lookback model relation described by eq. 5 and 6, whereas the dashed
and dashed-dotted lines use a shift of log A$_0$ in eq. 6 of this
relation by $\pm$ 0.5$\cdot$0.375. See text.    
  \label{Fig20} }
 \end{center}
\end{figure}

Fig.~\ref{Fig20} shows again the crucial relationship between gas mass
and stellar  mass for the Magneticum snap shot galaxies, however this time we
have introduced a color code with respect to the SFR.  Galaxies with log
$\delta$ = $\left|\psi/\psi_{LB}\right| \leq$ 0.20 are plotted as
green circles, galaxies with $\delta \geq$ 0.20 in blue and galaxies
with $\delta \leq$ 0.20 in yellow. We see a clear offset of the blue
and yellow circles relative to the green cicles in the overplotted
background. This means that the Magneticum simulations also contain an SFR
dependence in the relationship between gas mass and stellar mass. While we
show only the highest redshifts of the snap shot sample, we note that
the effect is also clearly present at lower redshift down to z =
0.5. Only at z = 0.1 the offset with SFR seems to disappear.

\begin{figure}[ht!]
 \begin{center}
   \includegraphics[width=0.45\textwidth]{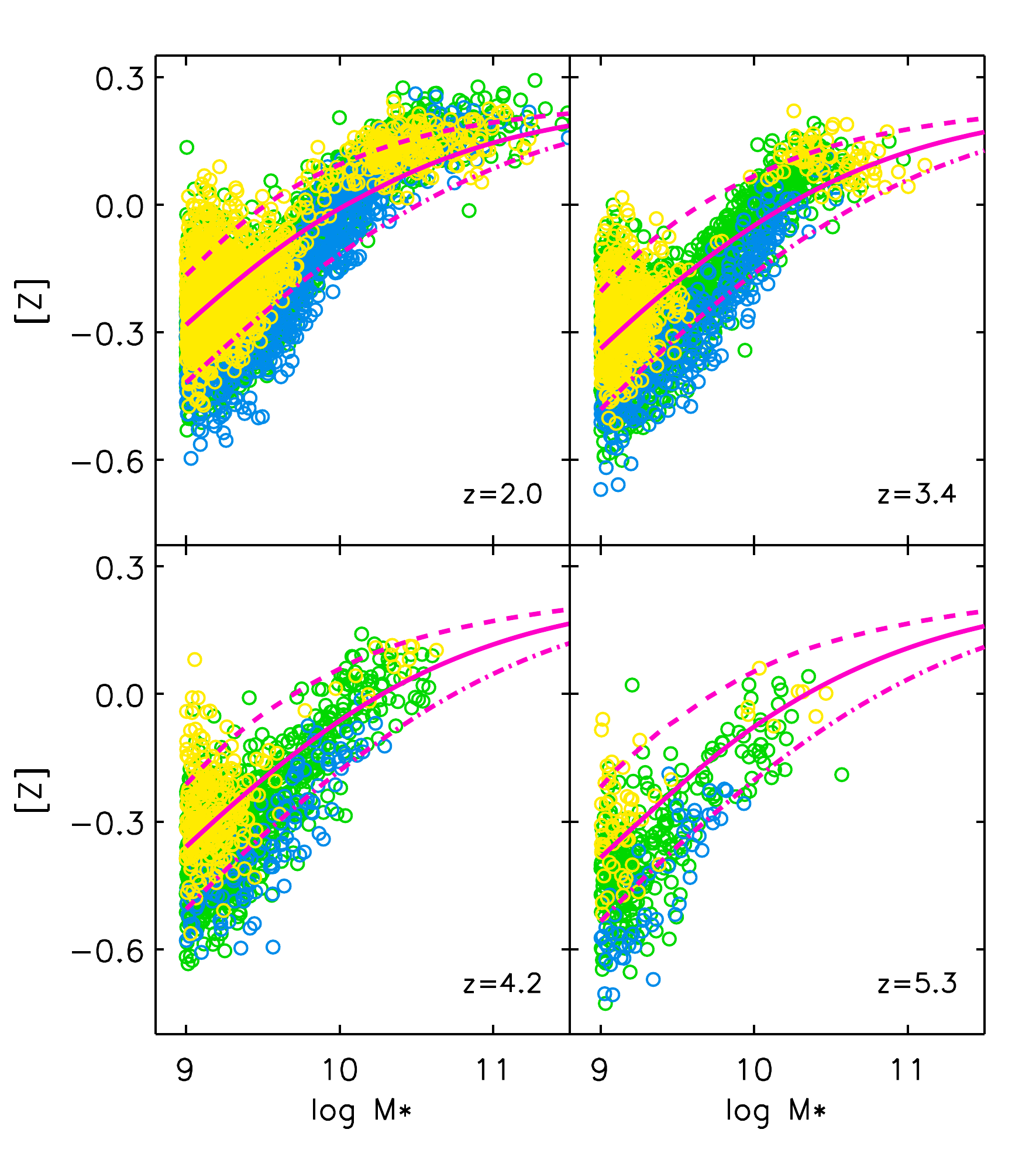}
   \caption{
Stellar metallicities of Magneticum galaxies versus stellar mass at four redshifts. The
galaxies are color coded according to their SFRs as in
Fig.~\ref{Fig20}. The pink solid curve  corresponds to lookback model
calculations with the main sequence star formation law of eq. 17 and
Appendix B adopted. The dashed and dashed-dotted curves apply shifts
in SFR and in the gas mass stellar mass relation as described in the text.
  \label{Fig21} }
 \end{center}
\end{figure}

We also include the standard relationship for the lookback models in
Fig.~\ref{Fig20} and for two models with $\Delta$log A$_0$ calculated
with $\delta$ = $\pm$0.375 and x$_{\delta}$ = $\pm$0.5,
respectively, as described above. We see that in this way the lookback
models cover the SFR dependent range of gas masses encountered in the
Magneticum simulations.

In Fig.~\ref{Fig21} we display the SFR color coded Magneticum
MZRs at the same four redshifts as in Fig.~\ref{Fig20} together with
the lookback models calculated with $\delta$ = 0, $\pm$0.375 and
x$_{\delta}$ = 0, $\pm$0.5 in the same way as described for
Fig.~\ref{Fig20}. The SFR dependence is clearly present in the
Magneticum snap shot sample and the lookback models capture this
effect well.

\section{Summary and discussion}

The main intention of the work presented here has been to develop lookback
galaxy evolution models as a simple tool to describe galaxy formation
and evolution, which can then be used  to interprete observational results derived from spectroscopy
such as the mass metallicity relationship or to calculate model
spectra using population synthesis techniques. As a crucial test of
this new tool we compare with the Magneticum cosmological simulations,
which describe the process of galaxy formation and evolution in a much
more comprehensive way.

An important ingredient for this comparison is the global galactic
SFR as a function of stellar mass and redshift. A whole variety of
such 'main sequence' relationships
derived from observations is available in the literature and we have
compared those with the Magneticum SFR. We found that the
Magneticum SFRs are in the right ballpark but there is a systematic
difference as a function of redshift. In order to use an SFR law for
our lookback models that represents the Magneticum calculations
well we have used the \citet{Pearson2018}, Appendix C, relationship but
with a correction factor as a function of redshift and with a
modification of the power law, which describes the dependence of
stellar mass.

In summary, we find that the lookback models capture many of the
Magneticum galaxy properties reasonably well. Most importantly, their
key assumption of a redshift dependent power law relationship between
ISM gas mass and stellar mass agrees well with the Magneticum
results. While a complex interplay of star formation, gas accretion
and mass-loss by galactic winds obviously affects the galactic gas
content, the net result, in a statiscal sense, is still that ISM gas
mass is related to the stellar mass at all redshifts. This has
important repercussions for the effective galactic gas accretion
rates, which are larger than the rates of mass-loss through galactic
winds but of the same order of magnitude as star formation rates.

The fraction of the cold star forming ISM gas to the
total gas changes continuously with time in the
Magneticum galaxies. At the early stage of galaxy formation all gas is
involved in the star forming process but then during the further
evolution with time the fraction of the gas contributing to star
formation becomes significantly smaller. This is an important factor
contributing to the main sequence star formation law. The specific star
formation rates of star forming galaxies strongly decrease during the course
of their evolution not because they are running out of gas (as
described by the lookback model power law relationship between gas
mass, redshift and stellar mass), but
because the fraction of the still present cold ISM gas, which is
capable of producing stars, becomes significantly smaller.

The luminosity weighted ages of the stellar population in the
Magneticum galaxies are also described reasonably well by the lookback
models. At low redshift there is a spread in ages, which can be explained by
the spread in SFR, which we encounter in the Magneticum galaxies.

With the assumption of a redshift dependent relationship between ISM gas
mass and stellar mass the lookback models predict that the average
metallicity of a star forming galaxy depends on the ratio of stellar
mass to ISM gas mass. Metal poor galaxies are gas rich and metal rich
galaxies are gas poor. Indeed, the Magneticum galaxies follow this
trend and the mass metallicity relationships at
different redshifts of Magneticum and the lookback models are in good
agreement. However, this requires an adjustment of the effective yield
in the lookback models, for which the yield was originally calibrated
so that observed MZRs in the local Universe obtained from spectroscopy
of young stars or the integrated stellar populations agreed with the
models. A comparison of Magneticum galaxies with these observations at
low redshift shows a small offset in metallicity which is then covered by
the adjustment of the effective yield.

The observed SFR rate dependence of the MZRs, which is also present in
the Magneticum galaxies, can be reproduced by the lookback models in a
very natural way. Galaxies with higher star formation rates are those with
higher gas mass content, because they also have more star forming gas
contributing to the star formation process. Therefore, their ratio of
gas mass to stellar mass is higher and, as metallicity in the lookback
models depends on the ratio of stellar to gas mass, their metallicity
is lower. 

The good agreement with the Magneticum
simulations confirms that the lookback models provide a simple
straightforward way to understand key aspects of the evolution of
galaxies, such as the average gas accretion history, the star
formation history and the formation of metals. This gives these models
a great potential as tool for population synthesis calculations of
synthetic spectra of the integrated stellar population of galaxies,
which can then be used to constrain stellar metallicities from
observed spectra. A typical example is given in the work by
\citet{Zahid2017}.

The advantage of the lookback models is that they are simple and easy
to calculate. Their chemical evolution can be described by a simple
analytical formula (see Appendix A) and a comparison with observations is
straightforward. However, important aspects have not yet been covered
in this first investigation. For instance, the chemical evolution
describes only metallicity as a whole and does not distinguish between
$\alpha$- and iron group elements. The inspection of the
Magneticum results shows that this approximation does not seem to lead
to large errors, because the ratio of $\alpha$ over iron abundances does
not change by more than 0.15 to 0.2 dex as a function of stellar mass
or redshift, but it is a systematic and expected trend and it would be
a clear improvement, if the models described this as well. We plan to implement
this in future work in a way that still keeps the simplicity of these models.

\acknowledgments
We thank our referee for an engaged constructive critical report.
This work has been supported by the Munich Excellence Cluster Origins
funded by the Deutsche Forschungsgemeinschaft (DFG, German Research
Foundation) under Germany's Excellence Strategy EXC-2094
390783311. The Magneticum Pathfinder simulations were
performed at the Leibniz-Rechenzentrum with CPU time assigned to
the Project ``pr86re''. We are especially grateful for the support by M.
Petkova through the Computational Center for Particle and Astrophysics
(C2PAP). \\

\appendix
\section{Analytical solution of the metallicity equation}

\begin{figure}[ht!]
 \begin{center}
   \includegraphics[width=0.40\textwidth]{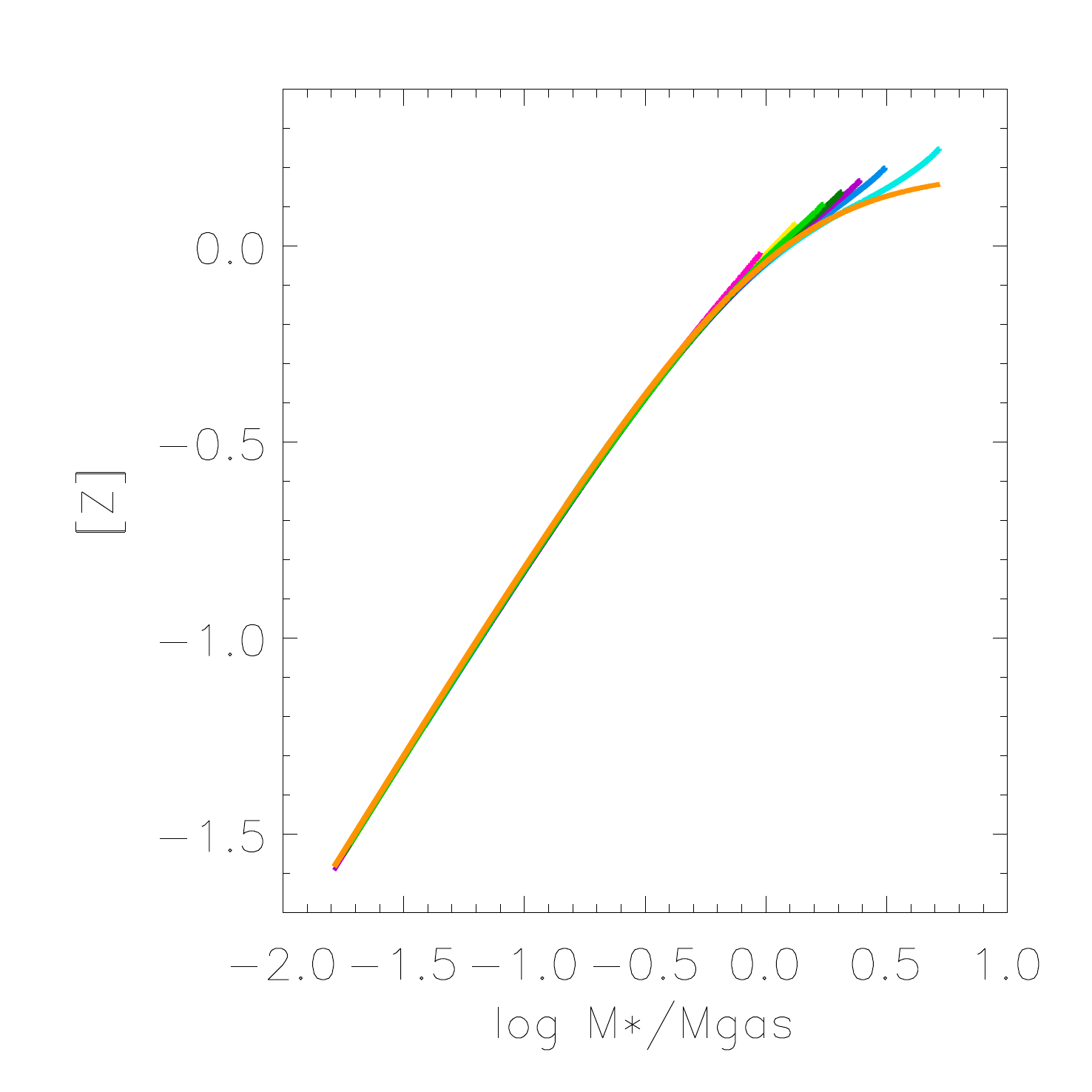}
   \includegraphics[width=0.40\textwidth]{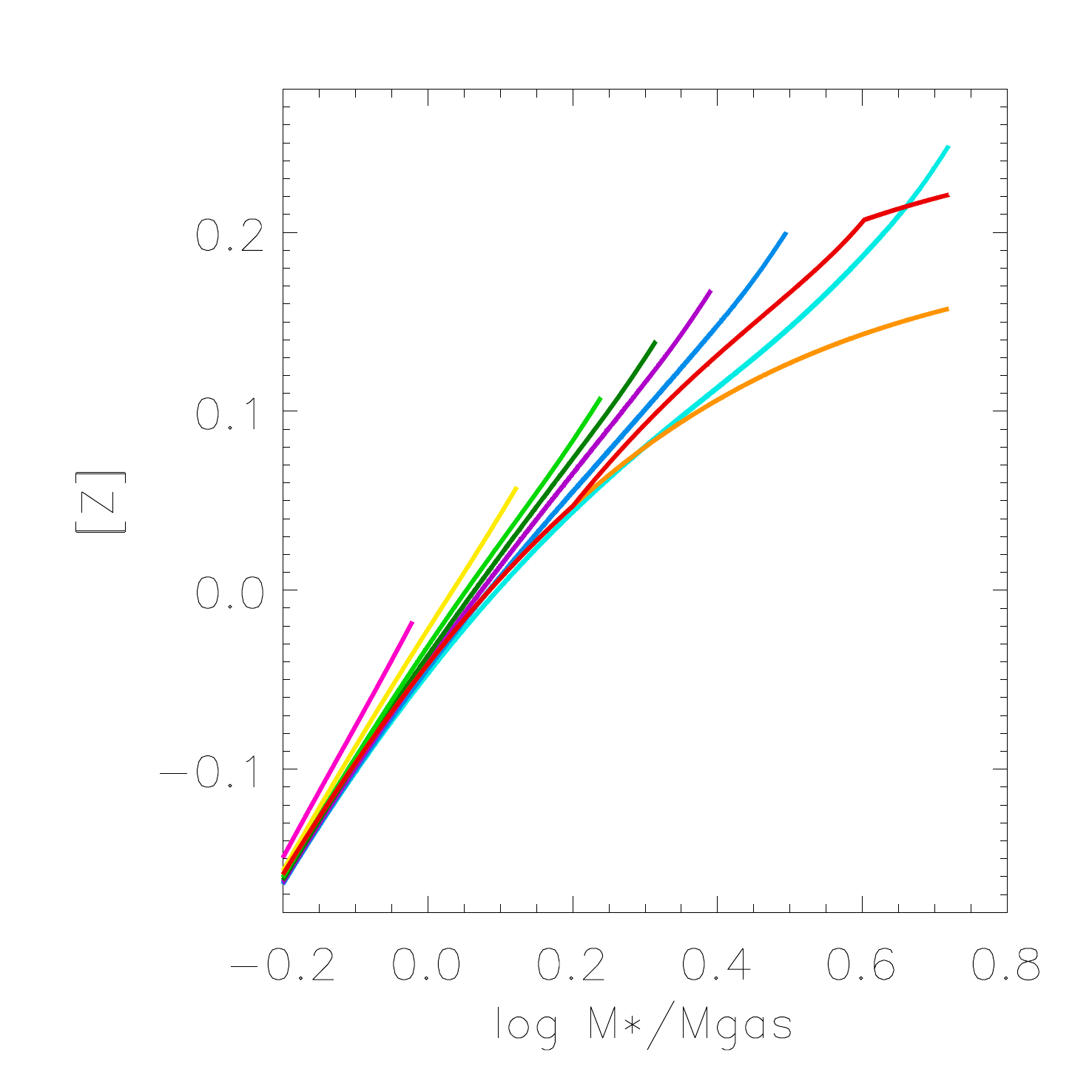}
  \caption{
Top: Same as Fig.~\ref{Fig1} but with the analytical solution of eq. (A6)
overplotted in orange. Bottom: Enlarged plot of the upper right corner
of the upper figure. The extended analytical solution of eq. (A8) is
added in red. \label{FigA1} }
 \end{center}
\end{figure}

Introducing the variable

\begin{equation}
x = {1 \over 1 - \beta(1-K)}{M_* \over M_g}
\end{equation}

eq. (13) turns into

\begin{equation}
{dZ \over dx} = {Y_N \over 1-R} -Z(1+{\mu \over x})
\end{equation}  

with

\begin{equation}
\mu = {\beta(1 - K) \over 1 - \beta(1-K)}.
\end{equation}  

The factor (1 - K) is roughly constant and close to unity for a large part of the evolution of a galaxy in our lookback model
approach except at the lowest redshifts and the largest ratios of
stellar to gas mass (log M$_*$/M$_g$ $>$ 0.3).  It is straightforward to show that under these conditions

\begin{equation}
Z \approx Y_N{1 \over 1 + \mu}x(1 - {1 \over 2 +
  \mu}x),~~~\mathrm{for}~ x \ll 1
\end{equation}  

or, alternatively,

\begin{equation}
Z \approx {Y_N \over 1-R}{M_* \over M_g}(1 - {1 \over 2 - \beta(1 - K)}{M_*
  \over M_g}),~~~\mathrm{for}~ x \ll 1.
\end{equation}  

Eq. (A4) and (A5) explain - at least for small values of x - why the
metallicity in our lookback models during the evolution of a galaxy
depends mostly on the ratio of stellar mass to gas mass M$_*$/M$_g$.

The analytical solution of eq. (A2) for $\mu$ = 1 is

\begin{equation}
Z = {Y_N \over 1-R}{1 \over x}(x-1+e^{-x}),~~~\mathrm{for}~ \mu = 1.
\end{equation}  

In Fig.~\ref{FigA1} we overplot this analytical solution (assuming 1 - K = 0.85) to the
individual numerical solutions of Fig.~\ref{Fig1}. We see that it is a
good match to the numerical solutions except for large M$_*$/M$_g$,
where it saturates and is $\sim$ 0.1 dex too small. We also see that
at these large values of stellar mass
the individual numerical solutions do not lie exactly on top of each
other but show a small dependence on the final mass at the
lowest redshift. The reason is that towards large M$_*$/M$_g$ values
(corresponding to low redshifts) the factor (1 -K) starts to change and becomes very
small and finally negative. This is caused by the increase of stellar
mass and the fact the star formation rate $\psi$ decreases towards
lower redshifts (see eq. 12 for the definition of K). As a consequence, the
term corresponding to dM$_g$/dM$_*$ on the right hand side of eq. (13)
becomes negative, which leads to an increase of Z rather than to a
saturation. The slight divergence of the numerical solutions for
different final stellar masses is the result of slightly different (1
- K) values in each model.

The behavior of K(M$_*$, z) at low redshift and large stellar masses
also means that our analytical approximation using the variable
transformation of eq. (A2) with (1 - K) = 0.85 breaks down. This
contributes to the discrepancy encountered in Fig.~\ref{FigA1}. However, if we
want to recover an analytical solution, we can introduce a correction
term, which changes the metallicity saturation behavior for large mass ratios log M$_*$/M$_g$ $\geqq$ 0.2

\begin{equation}
  c_Z(x)={1 \over 5}({\mu (x) \over x} - {\mu (x_0) \over x_0}),
  ~~~\mathrm{for}~ x \geqq x_0
\end{equation}

with x$_0$ the x-value at log M$_*$/M$_g$ = 0.2. We limit c$_Z$(x) to a
minimum value of -0.075 at the largest x-values. With this correction
our new analytical solution is

\begin{equation}
Z = {Y_N \over 1-R + c_Z(x)}{1 \over x}(x-1+e^{-x}),~~~\mathrm{for}~x \geqq x_0 ,
\end{equation}  

which is shown in the enlarged plot on the right hand side
of Fig.~\ref{FigA1}. In this plot we can again see how the individual
numerical solutions for different final stellar masses diverge for log
M$_*$/M$_g$ $\gtrapprox$ 0.2. As discussed above, the reason is the behavior of (1
- K) at large stellar masses, which is slightly different for each
numerical solution. However, since this is a small effect, our main
conclusion that in our look back models Z is a function of mainly
M$_*$/M$_g$ remains valid.
\section{Formulae for $\psi_{LB}$, x$_g^{fit}$ and $\tau_{fit}$}

As described in section 5 we use the \citet{Pearson2018}
SFR law of their Appendix C with several modifications. For the zero
point we introduce the correction factor c(z) as described by eq. (17).
In addition, we introduce a broken power law with respect
 to stellar mass
\begin{equation}
  \psi_{LB} = \psi_0^{LB}(z)(M_*/10^{10.5})^{\delta(z)}, M_* \ge 10^{9.8},
 \end{equation}

 \begin{equation}
  \psi_{LB} = 0.2^{\delta(z)}\psi_0^{LB}(z)(M_*/10^{9.8})^{1.1}, M_* \le 10^{9.8}.
 \end{equation}

$\delta$(z) and $\psi_0(z)$ are given in \citet{Pearson2018} in their
Appendix C.

For redshifts z $\ge$ 3.42 we need to introduce an additional
modfication to match the Magneticum SFRs at very low masses. We apply eq. (19) only in the range
10$^{9} \le $M$_* \le$ 10$^{9.8}$. For lower masses we use

\begin{equation}
  \psi_{LB} =
  0.2^{\delta(z)}0.132\psi_0^{LB}(z)(M_*/10^{9})^{\delta_l}, M_* \le 10^{9} 
\end{equation}

with

\begin{equation}
\delta_l(z) = 1.1 -0.6(1 - e^{({z - 3.42 \over 0.4})^2}).
 \end{equation} 

 The introduction of $\delta_l(z)$ leads to a transition of the power law
 exponent in this lower mass range from 1.1 to 0.5 when redshift
 increases. We find this trend in the low mass galaxies of the
 Magneticum evolution sample. 

Our fit of star forming to total mass of the cold ISM gas, x$_g^{fit}$
is calculated by

\begin{equation}
x_g^{fit} = 1 - a_g \cdot e^{-z/2}
\end{equation}  

with

\begin{equation}
a_g(m) = 0.89 + 0.08 \cdot e^{4(m_{min}-m)}, m=\mathrm{log}M_*, m_{min} = 9.00 - 0.3(1 - e^{-5z}).
\end{equation}

The factor e$^{-z/2}$ shifts x$_g^{fit}$ upwards with increasing
redshift z, until it saturates at unity. The second term in a$_g$
leads to a drop at low log M$_*$ and m$_{min}$  regulates at which
mass the drop sets in as function of z.

For z $\ge$ 1.5 and  log M$_* \ge$ 9.6 we need to account for the
redshift dependent decline of x$_g$ with stellar mass. We accomplish
this by introducing $\Delta$z = z - 1.5 and using

\begin{equation}
x_g^{fit} = (1 - a_g(9.6) \cdot e^{-z/2}) 10^{-p(\Delta z)(m - 9.6)}, z
    \ge 1.5, m \ge 9.6
  \end{equation}

  with

  \begin{equation}
p(\Delta z) = 0.02\Delta z(1+1.1\Delta z - 0.11\Delta z^2), a_g(9.6) =
0.89 + 0.08 \cdot e^{4(m_{min}-9.6)}.
\end{equation}

The Magneticum fit for the star formation time is

\begin{equation}
  \mathrm{log}~\tau_{fit} = a_{\tau}(z) -0.11 \cdot z(m-10.7), a_{\tau} = a_0(z) -0.25\cdot z
\end{equation}

with

\begin{equation}
  a_0 = 9.15 +0.22\cdot w(z), w(z) = {e^{-3(z-2.8)} \over 1 + e^{-3(z-2.8)}}.
 \end{equation}

The function w(z) causes a switch from a$_0$ = 9.37 at low z to a$_0$ =
9.15 at high z.

\section{Calculation of luminosity weighted averages}

In order to calculate the V-band luminosity weighted ages and
    stellar metallicities of the stellar population of our lookback
    models we make the simplifying assumption to consider only the
    contribution by hydrogen burning main sequence stars, since they
    provide the by far strongest contribution to the integrated light
    at V-band. Along the stellar main sequence the V-band luminosity
    at stellar mass m is

\begin{equation}
L_V (m) = {L \over L_{\odot}}(m) 10^{0.4(BC(m) - BC_{\odot})},
\end{equation} 

where L(m)/L${\odot}$ is the mass-luminosity relation and
BC(m) is the bolometric correction as a function of stellar main
sequence mass. The V-band luminosity of a population at age t is then

  \begin{equation}
  L^{pop}_V (t) = \int_{m_{min}}^{m_{max}(t)} N(m) L_V (m) dm.
  \end{equation}

The upper limit of the integral m$_{max}$(t) is defined by the mass
dependent stellar lifetime on the main sequence $\tau_{m}$ and given
by the mass at which $\tau_{m}$ = t. N(m) is the
\citet{Chabrier2003} initial mass-function. The maximum value of
m$_{max}$ is 100 M$_{\odot}$ and m$_{min}$ is adopted as
0.07 M$_{\odot}$. The main sequence lifetime-mass, mass-luminosity,
effective temperature-mass relations and the bolometric
corrections are obtained from \citet{Ekstrom2012},
\citet{Kaltenegger2009}, \citet{Southworth2015} (the DEBcat catalogue),
\citet{Flower1996}, and \citet{Torres2010}, respectively.

The luminosity contribution of the stellar population of age t to the
  total galaxy V-band luminosity L$^{tot}_V (\tau)$ at a time $\tau
  \geq$ t then depends on how many
  stars have formed at t and is, therefore, given by

\begin{equation}
W_V (t) = {1 \over L^{tot}_V} L^{pop}_V (t) \psi (t) dt
\end{equation}

with

\begin{equation}
L^{tot}_V (\tau) = \int^\tau_0 W_V (t) dt.
\end{equation}

The stellar metallicity [Z]$_* (\tau)$ is then given by the
  V-band luminosity weighted avarage of the metallicities of the stars that
  formed at earlier times

\begin{equation}
Z_* (\tau) = \int^\tau_0 [Z] W_V (t) dt.
\end{equation}

In the same way, the luminosity weighted age  a($\tau$) of the stellar
  population of a galaxy at time $\tau$ is calculated as

\begin{equation}
a (\tau) = \int^\tau_0 t W_V (t) dt.
\end{equation}

\bibliography{ms}

\end{document}